\pdfoutput=1
\documentclass{elsart}

\usepackage{natbib}

\usepackage{amsmath}
\usepackage{amssymb}
\usepackage{graphicx}
\usepackage{verbatim}
\usepackage{rotating}
\usepackage{multirow}

\DeclareGraphicsExtensions{.pdf}

\begin{document}

\begin{frontmatter}



 \title{Dynamics of Unperturbed and Noisy Generalized Boolean Networks}

\author[addr1,addr2]{Ch. Darabos\thanksref{thx1}}
\ead{Christian.Darabos@unil.ch}
\ead[url]{www.hec.unil.ch/isi}
\author[addr1]{M. Tomassini\thanksref{thx1}}
\ead{Marco.Tomassini@unil.ch}
\author[addr2,addr3]{M. Giacobini\thanksref{thx2}}
\ead{Mario.Giacobini@unito.it}

\thanks[thx1]{Partially supported by grant 200021-107419/1 of the Swiss National Science Foundation.}
\thanks[thx2]{Partially funded by the Ministero dell'Universit\'a e della Ricerca Scientifica e Tecnologica}

\address[addr1]{Information Systems Institute\\
Faculty of Business and Economics - University of Lausanne, Switzerland}

\address[addr2]{
Computational Biology Unit\\
Molecular Biotechnology Center - University of Torino, Italy
}

\address[addr3]{
Department of Animal Production, Epidemiology and Ecology\\
Faculty of Veterinary Medicine - University of Torino, Italy
}

\begin{abstract}
For years, we have been building models of gene regulatory networks, where recent advances in molecular biology shed some light on new structural and dynamical properties of such highly complex systems. In this work, we propose a novel timing of updates in Random and Scale-Free Boolean Networks, inspired by recent findings in molecular biology. This update sequence is neither fully synchronous nor asynchronous, but rather takes into account the sequence in which genes affect each other. We have used both Kauffman's original model and Aldana's extension, which takes into account the structural properties about known parts of actual GRNs, where the degree distribution is right-skewed and long-tailed. The computer simulations of the dynamics of the new model compare favorably to the original ones and show biologically plausible results both in terms of attractors number and length. We have complemented this study with a complete analysis of our systems' stability under transient perturbations, which is one of biological networks defining attribute. Results are encouraging, as our model shows comparable and usually even better behavior than preceding ones without loosing Boolean networks attractive simplicity.
\end{abstract}


\end{frontmatter}

\section{Introduction}
\label{sect:intro}

Gene regulatory networks comprising genes, proteins and other interacting molecules,  are extremely complex systems and we are just beginning understanding them in detail.
However, it is possible, and useful, to abstract many details of the particular kinetic equations
in the cell and focus on the system-level properties of the whole network dynamics. This Complex Systems Biology approach,
although not strictly applicable to any given particular case, may still provide interesting general insight.\\
Random Boolean Networks (RBNs) have been introduced by Kauffman more than thirty years ago~ \cite{kauffman69} as a highly simplified model of gene regulatory networks (GRNs). RBNs have been studied in detail by 
analysis and by computer simulations of statistical ensembles of networks and they have been shown to be capable of surprising
dynamical behavior. We summarize the main results in the next section.\\
In the last decade, a host of new findings and the increased availability of biological data have changed
our understanding of the structure and functioning of GRNs. In spite of this, the original view of Kauffman has been used to predict gene expression patterns observed experimentally ~\cite{bilke2001, alvarezbuylla2008}. Today, this model is still valid provided that it is updated to take into account the new knowledge about the topological structure and the
 timing of events of real-life
 gene regulatory networks without loosing its attractive simplicity. Following these guidelines, our aim
 in this work is to describe and test a new model that we call Generalized Boolean Networks (GBNs), 
 which includes, at a high level of abstraction, structures and mechanisms that are hopefully closer
 to the observed data.
 Adhering to the original Kauffman's view that attractors of the dynamics of RBNs are the important feature and that they roughly correspond to
 cell types, 
  we will discuss the results of the systems ability to relax into stable cycles or fixed points, and their tolerance to local perturbation.\\
The organization of this work is the following. In the next section we briefly review
the main assumption implied in Kauffman's RBNs and their possible limitations.  Changes to both
the randomness and the synchrony assumptions will be proposed in section~\ref{sect:gbn} leading
to generalized boolean networks. In sections~\ref{subsect:setting} and~\ref{subsect:analysis} the
new model is studied by statistical sampling using numerical simulation. 
Then we introduce the concept of perturbation in section~\ref{sect:fault} and we investigate numerically
the stability properties of GBNs.
Finally, in section ~\ref{sect:concl} we present our conclusions and discuss possible future work.

\section{Classical Random Boolean Networks}
\label{sect:rbns}

Random Boolean Networks (RBNs) have been introduced by Kauffman~ \cite{kauffman69} as a highly simplified model of gene regulatory networks.
In Kauffman's RBNs with $N$ nodes, a node represents a gene and is modeled as an on-off device,  meaning 
that a gene is expressed if it is on (1), and it is not otherwise (0). Each
gene receives $K$ randomly chosen inputs from other genes.
Initially, one of the $2^{{2}^{K}}$ possible Boolean functions of $K$ inputs  is assigned at random to each gene.
The network dynamics is discrete and synchronous: at each time step all nodes simultaneously
examine their inputs, evaluate their Boolean functions, and find themselves in their new states at the next time step.\\
More precisely, the \textit{local transition rule} $\phi$ is one of the $2^{{2}^{K+1}}$ possible Boolean functions of $K$ inputs from the neighboring nodes plus that of the node itself, thus possibly implementing a biological situation where a gene regulates itself:
\[
   \phi:\Sigma^{K+1} \rightarrow \Sigma.
\]
This function maps the state $s_{ i} \in \Sigma =\{0,1\}$ of a given node $i$ into
another state from the set $\Sigma$, as a function of the state of the node itself and of the states of the
nodes that send inputs to $i$. \\
For a finite-size system of size $N$ (such as those treated herein) a {
\textit{configuration}  $C(t)$ of the RBN at time $t$ is defined by the binary string:
\[
  C(t) = ( s_{0}(t), s_{1}(t), \ldots, s_{N-1}(t) ),
\]
where $s_{i}(t) \in \Sigma$ is the state of node $i$} at time
$t$. The progression of the RBN in time is then given by the iteration
of the {\textit{global mapping}, also called \textit{evolution operator} $\Phi$:
\[
 \Phi: C(t) \rightarrow C(t+1), \verb+   + t=0,1, \ldots
\]
through the simultaneous application at each node of the non-uniform local
transition rule $\phi$.  The global dynamics of the RBN can be described
as a directed graph, referred to as the RBN's {\textit{phase space}.
Over time, the system travels through its phase space, until a point or cyclic attractor is reached whence either it will remain in that point attractor forever, or 
it will cycle through the states of the periodic attractor. Since the system is finite and deterministic, this will happen at most after 
$2^N$ time steps.\\
This extremely simple and abstract model has been studied in detail by ana\-lysis and by computer
simulations of statistical ensembles of networks and it has been shown to be capable of surprising
dynamical behavior. Complete descriptions can be found in \cite{kauffman93,aldana-copp-kadanoff-03}. 
We summarize the main results here. \\
First of all, it has been found that, as some parameters are 
varied such as $K$, or the probability $p$ of expressing a gene, i.e. of switching on the corresponding 
node's state, the RBN can go through a phase transition. Indeed, for every value of $p$, there is a critical
value of connectivity:

$$K_c(p) = [2p(1-p)]^{-1}$$  

such that for values of $K$ below this critical value $K_c(p)$ the
system is in the ordered regime, while for values of $K$ above this limit the system is
said to be in the chaotic regime. In classical RBNs  $K_c(0.5) = 2$ corresponds to the edge  between the ordered and the chaotic regime,
systems where $K < 2$ are in the ordered regime, and $K > 2$ means that the system is in the chaotic phase for $p = 0.5$.\\
In his original work, Kauffman discovered that the mean cycle length scales are at most linear with $N$ for $K=2$. He also believed that the number of attractors scales with the square root of the number of genes in the system, which has an interesting analogy with the number of different cell types for genomes in multicellular organisms. In fact, this last hypothesis has been proven to be an artifact of undersampling by Bilke and Sjunnesson~\cite{bilke2001} who showed that the number of attractors scales linearly with $N$. In addition, Kauffman found that for $K=2$ the size distribution of perturbations in the networks is a power-law with finite cutoff that scales as the square root of $N$. Thus perturbations
remain localized and do not percolate through the system. Kauffman's suggestion was that cell types correspond 
to attractors in the RBN phase space, and only those attractors that are short (between one and a few tens or hundreds of states) and stable under
perturbations will be of biological interest. Thus, according to Kauffman, $K=2$ RBNs lying 
at the edge between the ordered phase and the chaotic phase can be seen as abstract models of 
genetic regulatory networks. RBNs are interesting in their own as complex dynamical systems and have been throughly 
studied as such using the concepts and tools of statistical mechanics (see
 \cite{derrida-86,aldana-copp-kadanoff-03}). \\
For the sake of completeness, let us mention that the ``discrete'' approach to the high-level
description of genetic regulatory networks is not the only possible one. A more realistic description
is obtained through the use of  a ``continuous-state'' model. In the latter, the levels of 
messenger RNA and proteins are assumed to be continuous functions of time instead of
on/off variables. The system evolution is thus represented by sets of differential
equations modeling the continuous variation of the components concentration. Here we focus
on the discrete approach, but the interested reader can find more information on the
continuous models in~\cite{glass-06}, for instance.\\


\section{From Random to Generalized Boolean Networks}
\label{sect:gbn}

In this section we describe and comment on the main assumptions implied in Kauffman's
RBNs. Following this, we propose some modifications that, in our opinion, should bring
the model closer to known facts about genetic regulatory networks, without loosing the
simplicity of classical RBNs.\\
\noindent Kauffman's RBN model rests on three main assumptions: 
\begin{enumerate}
\item Discreteness: the nodes implement Boolean functions and their state is either on or off;
\item Randomness: the nodes that affect a given node in the network are randomly chosen and
      are a fixed number;
\item Timing: the dynamics of the network is synchronous in time.
\end{enumerate}

\subsection{Discrete State Approach} The binary state simplification could seem extreme but actually it represents quite
well ``threshold phenomena'' in which variables of interest suddenly change their
state, such as neurons firing or genes being switched on or off. This can be understood
since the sigmoidal functions one finds in the continuous differential equation approach~\cite{glass-06}
actually do reduce to threshold gates in the limit, and it is well known that Boolean functions can
be constructed from one or more threshold gates~\cite{hassoun95}. So, in the interest of simplicity,
our choice is to keep the discrete Boolean model for the states of the nodes and the
functions implemented at each node.\\

\subsection{Random vs Scale-Free Networks}
\label{rndvssf}
RBNs are directed random networks. The edges have an orientation because they represent a chemical
influence from one gene to another, and the topologies of the graphs are random because any node is as likely to
be connected to any other node in an independent manner. There are two main types of RBNs, one
in which the connections are random but the degree of each node is fixed, and a more general one in which only
the average connectivity is fixed.
Random graphs with fixed connectivity degree were a logical generic choice in the
beginning, since the exact couplings in actual genetic regulatory networks were largely unknown. 
Today it is more 
open to criticism since it does not correspond to what we know about the topology of
biological networks. In fact, many biological networks, including genetic regulatory
networks, seem to have a scale-free type or hierarchical output distribution (see, for 
example,\cite{vazquez-topo-nets-04,albert05,albert07}) but not random, according to present data,
as far as the \textit{output} degree distribution is concerned~\footnote{The \textit{degree distribution function} $p(k)$ of a  graph represents the probability that a randomly
chosen node has degree $k$~\cite{newman-03}. For directed graphs, two distributions may be defined,
one for the outgoing edges $p_{out}(k)$ and another for the incoming edges $p_{in}(k)$.}.
The \textit{input} degree
distributions seem to be close to normal or exponential instead. A scale-free distribution for the degree means that 
$p(k)$ follows a power-law $p(k) \sim k^{- \gamma}$, with $\gamma$ usually but not always between $2$ and
$3$. In contrast, random graphs have a Poisson 
degree distribution $p(k) \simeq \bar k^k \, e^{-\bar k} /k!$, where $\bar k$ is the mean degree, or a delta distribution as in a classical fixed-degree RBN. Thus the low fixed connectivity
suggested by Kauffman ($K \sim 2$) for candidate stable systems is not found in such degree-heterogeneous networks, where a  wide connectivity range is observed instead. 
The consequences for the dynamics may be important, since in scale-free graphs there are many nodes with  low degree and a small, but not vanishing, number of highly connected nodes (see, for instance,
~\cite{alb-baraba-02,newman-03}).\\
For the sake of completeness, we also wish to point out that the degree distribution is only one statistical aspect of
a given network and the attribution of a scale-free nature to some genetic regulatory networks has
been challenged~\cite{dupuy-et-al,tanaka}. Indeed, it has been recently shown that a random sample of networks with different degree
distributions may give subgraphs with similar degree distributions. Conversely, networks with
identical degree distributions may have different topologies~\cite{hormoz,dupuy-et-al}. 
The issue is still far from being settled due to the insufficient amount of analyses. However, we believe that it doesn't
fundamentally change the nature of high-level models such as those discussed here. In particular,
everybody seems to agree on the fact that the distributions are, if not scale-free, at least broad-scale, i.e
they have a longer tail to the right for the output degree distribution.\\
The first work that we are aware of, using the scale-free topology for modeling Boolean networks 
dynamics is \cite{oosawa-sav-02}. Oosawa and Savageau took \textit{Escherichia coli} as a model
for their scale-free nets with an average input degree $\bar k$ of two. Interesting 
in this particular case, the model is a little too specialized as most other known networks or network fragments have higher connectivity levels. 
What is needed are models that span the range of observed connectivities. \\
Along this line, Aldana presented the first detailed analysis of a model Boolean network with scale-free
topology \cite{aldana-sf-03,aldana-cluzel}. Using the power-law exponent $\gamma$ as a critical parameter instead of the
mean degree, he has been able to define a phase space diagram for scale-free
boolean networks, including the phase transition from ordered to chaotic dynamics, as a function of 
$\gamma$ where, if $p=0.5$, then $\gamma_c(p) = 2.5$ is the critical value for which systems rest on 
the edge between order and chaos, if $\gamma_c(p) > 2.5$ and the system is in the ordered regime and if $\gamma_c(p) < 2.5$ it lies in the chaotic phase. He also made exhaustive simulations  for several small values 
of the network size $N$ ($N \le 20$). The scale-free distribution was the input distribution $p_{in}(k)$ while
$p_{out}(k)$ was Poissonian.  We now know these distributions are actually inverted when compared to known GRNs. In our model we have thus adopted networks 
with a scale-free output distribution, and a Poissonian input distribution, as this seems to be at least close to the actual topologies.  However, from the mathematical point of view the results in terms of
different regimes as a function of $\gamma$ are the same in both cases~\cite{aldana-sf-03,aldana-cluzel}.\\
One problem with Aldana's networks was their small size since he wanted to explore the phase space
exhaustively, and this can only be done for small $N$. However, scale-free network statistics cannot
be accurate unless the network size is large enough and $k$ ranges, which should span at least
a few orders of magnitude, are suitably binned or the
cumulative distribution function is used instead of $p(k)$~\cite{dor-mendes}.
In another recent work, RBNs of various topologies and of larger size have been studied using
statistical sampling and numerical simulation by Iguchi et al.~\cite{iguchi07}. They used 
standard synchronous 
updating of network nodes and various graph topologies: random with Poisson distribution,
exponential, and scale-free. Iguchi et. al focused on the distribution of phase space attractors
and on their lengths and as such their work is closely related to the one presented here. However, 
most of their results concern the directed networks in which the input and
output distributions are the same ($p_{in}(k)=p_{out}(k)$) and, as said, above the timing of node update is synchronous. 
They focused their analysis on the mean degree
$\bar k$. While $\bar k$ is a significant parameter for random and exponential degree-distributed
graphs, it is much less meaningful for graphs having a scale-free degree distribution. For a continuous
 power-law distribution defined in $(0,+\infty)$ the mean becomes infinite for $\gamma \le 2$ and the variance diverges for $\gamma \le 3$~\cite{sornette}. Although
$\bar k$ can always be computed given a finite arbitrary degree 
sequence $\{k_j\}, j=1,\ldots,N$, it still
looses its meaning when the distribution is such that a non-negligible number of extreme
values exist, as in scale-free networks which are highly degree-heterogeneous. In this case, the
average is controlled by the few largest degrees and not by the numerous small ones.
These differences make it difficult to directly compare their results
with ours but we shall nevertheless comment on our respective findings as their study is related in
many ways to the present one.

\subsubsection{Construction of input and output degree networks distributions}
\label{nets}
Here we present the methodology for constructing our model networks, starting with the input and output degree distributions.
As said above, Kauffman's RBNs are directed graphs with connectivity $K$. In fact, as anticipated in the preceding section, according to present data many biological networks, including GRNs, suggest an 
inhomogeneous output distribution and a Poissonian or exponential input distribution~\cite{vazquez-topo-nets-04,albert07}. Whether $p_{in}(k)$ is Poissonian or exponential  both distributions have a tail that decays quickly, although the Poissonian distribution does so
even faster than the exponential, and thus both have a clear scale for the degree. On the other hand, $p_{out}(k)$ is very different, with a fat
tail to the right, meaning that there are some nodes in the network that influence many other
nodes. \\
In our model we have thus adopted networks with a scale-free output distribution, where $p_{out}(k)$ is the probability a node $n$ will have a degree $k$:
$$p_{out}(k) = \frac{1}{Z} k^{-\gamma} $$ 
where the normalization constant $Z(\gamma) = \sum_{k=1}^{k_{max}} k^{-\gamma}$ coincides with Riemann's Zeta function for $k_{max} \rightarrow \infty$. The input distribution approximates a normal function centered 
around $\bar k$. We call our model scale-free boolean networks (SFBNs). Figure \ref{degree_distr} offers a taste of what such distributions look like. 
\begin{figure*} [!ht]
\begin{center}
\begin{tabular}{cc}
\mbox{\includegraphics[width=6.5cm]{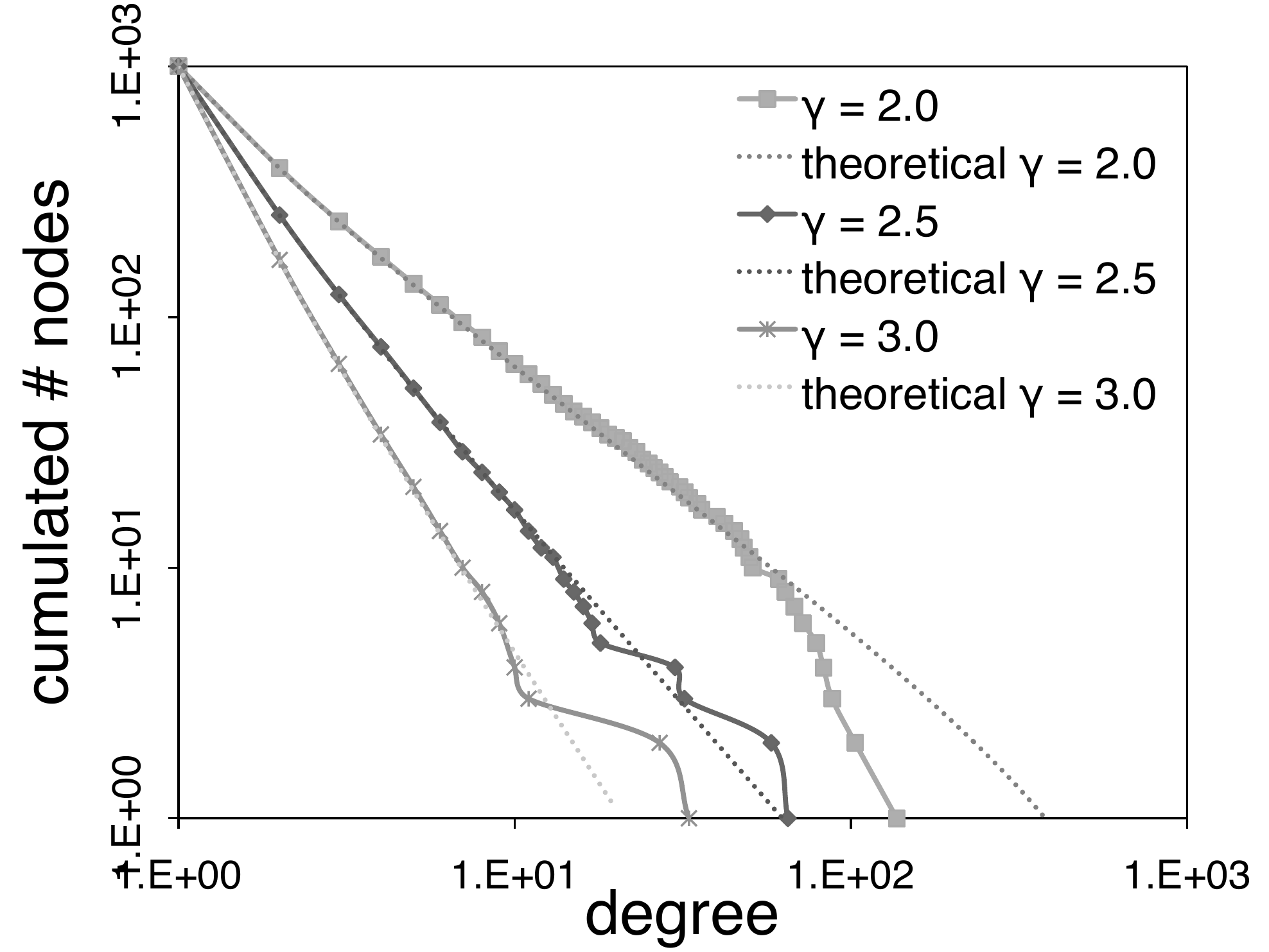}} \protect &  		
\mbox{\includegraphics[width=6.5cm]{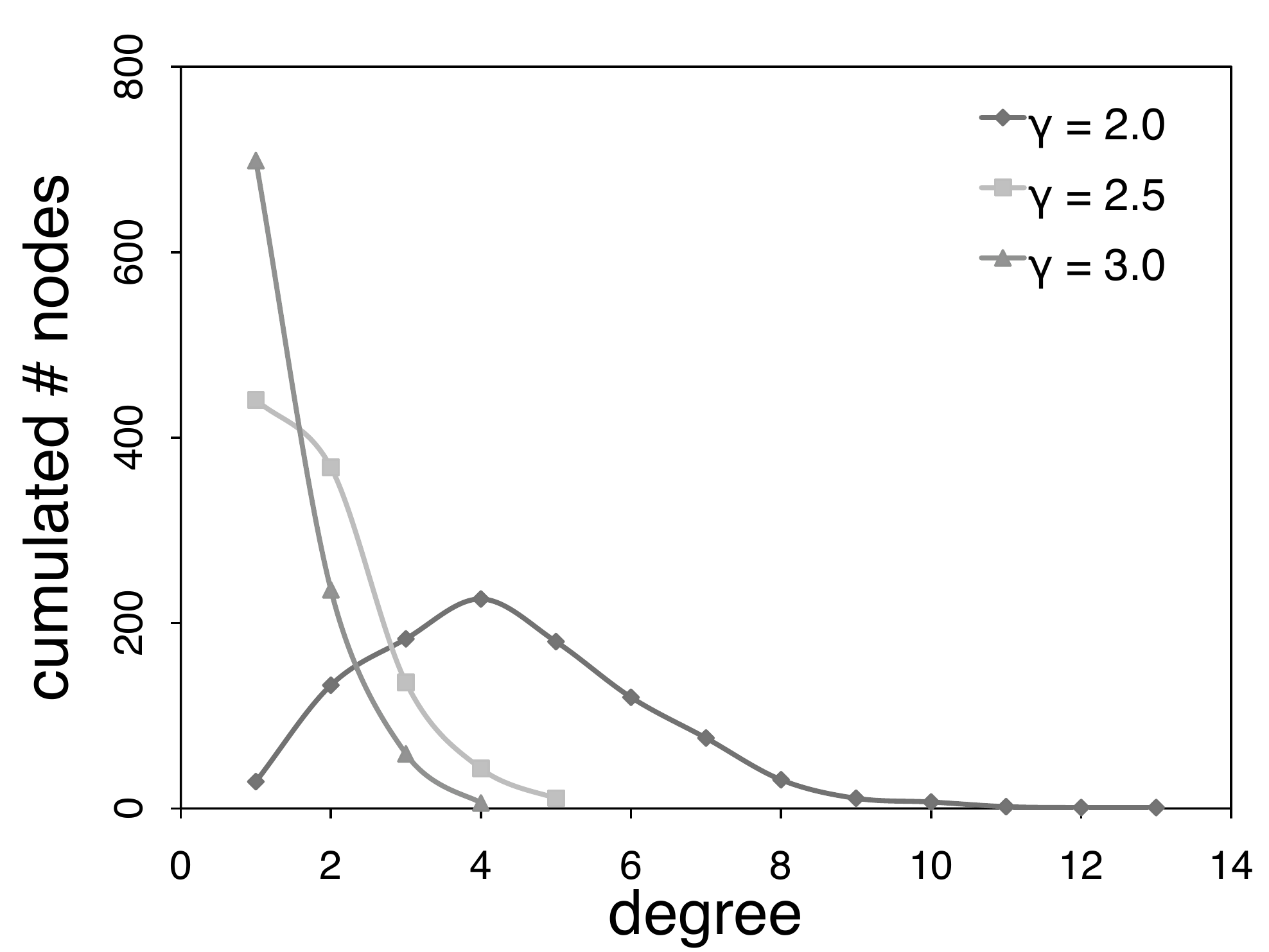}} \protect\\
(a)&(b)\\
\end{tabular}
\end{center}
\caption{Networks degrees distributions. Actual output degrees distributions on a log-log scale (a)  and input degrees distributions on a lin-lin scale (b) of a sample generated networks of size $N = 1000$ and {$\gamma = 2.0$}, {$\gamma = 2.5$} and, {$\gamma = 3.0$}. Distributions are discrete and finite; the continuous lines are just a guide for the eye.}
\label{degree_distr}
\end{figure*}
Naturally, $p(k)$ being defined over the positive integers only, an approximation is necessary to define how many nodes of the network have a given input degree. Namely, in a first pass, we use the integer value $\lfloor p(k)\rfloor$ as the number of nodes that will have a degree $k$ , for each degree $k \in \{k_{min}, k_{max}\}$. In a second pass, we use the decimal value $p(k)-\lfloor p(k)\rfloor$ as the probability that one more node will have a degree $k$ until the degree of all nodes has been specified. This non-deterministic process causes slight differences in the distributions, which is especially important around the critical regime to explore the solution space since the previously mentioned approximation leaves each scale-free distributions slightly off the power-law. Once the exact output degree distribution of a given network is known, we use the average connectivity $\bar k$ to produce a matching discrete Poisson input distribution. Finally, each node $i$ of the system is assigned an input degree $k^i_{in}$ and an output degree $k^i_{out}$, and nodes are randomly connected according to these, avoiding edge repetitions.

In Table \ref{table_avg_k} below, we show the average input and output degrees $\bar k$ over 100 different networks, including their standard deviation. This is only given as an information, because, as it has been mentioned in section \ref{rndvssf} of this work and according to Aldana's model \cite{aldana-sf-03}, the regime of a SFBN cannot be defined by its average connectivity but by the $\gamma$ exponent of its output degree distribution function.

\begin{table*}[!ht]
\begin{center}
\begin{tabular}{c|c|c|c|}
& $\bar{k}_{order}$ & $\bar{k}_{critical}$ & $\bar{k}_{chaos}$ \\
\hline
$N=100$ & $ 1.36_{\pm 0.06}$ & $ 1.82_{\pm 0.14}$ & $3.25_{\pm 0.35}$ \\
$N=150$ & $ 1.39_{\pm 0.05}$  & $ 1.81_{\pm 0.11}$ & $ 3.42_{\pm 0.42}$ \\
$N=200$ & $ 1.38_{\pm 0.04}$  & $ 1.81_{\pm 0.11}$ & $ 3.68_{\pm 0.40}$ \\
$N=500$ & $ 1.37_{\pm 0.02}$ & $ 1.84_{\pm 0.07}$ & $ 3.78_{\pm 0.29}$ \\
$N=750$ & $ 1.37_{\pm 0.02}$ & $ 1.84_{\pm 0.06}$ & $ 3.85_{\pm 0.24}$ \\
$N=1000$ & $ 1.37_{\pm 0.02}$ & $ 183_{\pm 0.05}$ & $ 3.98_{\pm 0.23}$ \\
\hline
\end{tabular}
\end{center}
\caption{Networks mean degrees. Average input and output degrees $\bar k$ over 100 networks including the standard deviation in all three different regimes.}
\label{table_avg_k}
\end{table*}

Iguchi et al.~\cite{iguchi07} have explored Boolean networks where both the output and the input distribution are of the scale-free type and used the average degree as an indicator of differentiation. Although an interesting metric, the average degree allows one to distinguish regimes only in Kauffman's classical RBNs with random topologies. Instead, they have used a modified {\em Barab\'asi-Albert preferential attachment} model which allows one to tune the networks average degree $\bar{k}$. However,  the $\bar{k}$ factor has no effect on the regime, as the preferential attachment model~\cite{alb-baraba-02} produces a single value of $\gamma \sim 3$ well into the chaotic regime. When dealing with the attractors cycle lengths, they have used systems where the average degree was either $\bar{k}=2$ or $\bar{k}=4$ which, according to our calculations, would essentially place all of their systems more or less deeply in the chaotic regime. In addition, they show examples of SFRBNs with an average degree of $\bar{k}=1$, which does not seem possible if all nodes are connected. Thus, a direct comparison
of our results with those of~\cite{iguchi07} is hardly meaningful for SFRBNs.


\subsection{Timing of Events}

Standard RBNs update their state synchronously (SU). This assumption simplifies the analysis, but
does not agree with  results  on gene activation experiments if the network has to be biologically plausible~\cite{glass-06}. Rather, genes seem
to be expressed in different parts of the network at different times, according to a strict
sequence (see, for instance, \cite{gen-net-02}). Thus a kind of serial, asynchronous
update sequence seems to be needed. Asynchronous dynamics must nevertheless be further qualified, since there 
are many ways for serially updating the nodes of the network. \\
Two types of asynchronous updates are commonly used. In the first, a random 
permutation (RPU) of the nodes is drawn and the nodes are updated one at a
time in that order.  At the next update cycle, a fresh permutation is drawn and the cycle is
repeated. In a second  often used policy, the next cell to be updated is chosen
at random with uniform probability and with replacement. This is a good approximation of 
a continuous-time Poisson process, and it will be called \textit{Uniform Update} (UU).\\
Several researchers have 
investigated the effect of asynchronous updating on classical RBN dynamics in recent years
\cite{harvey97,christof-ecal,Gershenson2004b}.
Harvey and Bossomayer
studied the effect of random asynchronous updating on some statistical properties of
network ensembles, such as cycle length and number of cycles, using both RPU and UU
 \cite{harvey97}. They found that many features that arise in synchronous RBN do
 not exist, or are different in non-deterministic asynchronous RBN. Thus, while point
 attractors do persist, there are no true cyclic attractors, only so-called ``loose'' ones
 and states can be in more than one basin of attraction. Therefore attractor lengths, which
 is one of the main features in RBNs, are not well defined in the asynchronous case.
Also, the average number of attractors is very different from the synchronous
 case: even for $K=2$ or $K=3$, which are the values that characterize systems at the
 edge of chaos, there is no correspondence between the two dynamics.\\
Mesot and Teuscher~\cite{christof-ecal} studied the critical behavior of asynchronous RBNs and concluded 
that they do not have a critical connectivity
value analogous to synchronous RBNs and they behave, in general, very differently from
the latter, thus confirming in another way the findings of \cite{harvey97}.\\   
Gershenson~\cite{Gershenson2004b} extended the analysis and simulation
of asynchronous RBNs by introducing
additional update policies in which specific groups of
nodes are updated deterministically. He found that all types of networks have the same
point attractors but other properties, such as the size of the attractor basins and the cyclic
attractors do change. \\
Considering the above results and what is known experimentally about the timing of events
in genetic networks we conclude, with~\cite{christof-ecal}, that neither fully synchronous
nor completely random asynchronous network dynamics are suitable models. Synchronous update is implausible
because events do not happen all at once, while completely random dynamics
does not agree with experimental data on gene activation sequences and the model does
not show stable cyclic attractors of the right size. For this reason, in the following section~\ref{semi-sync}
we propose a new quasi-synchronous node update scheme, which is closer to that observed in natural systems \cite{gen-net-02,olivieri-davidson-04}.


\subsubsection{Semi-Synchronous Update Scheme}
\label{semi-sync}

As we have seen above, in GRNs, the expression of a gene depends on some transcription factors, 
whose synthesis appears to be neither fully synchronous nor instantaneous. Moreover, 
in some cases like the gene regulatory network controlling embryonic specification in the sea urchin
\cite{gen-net-02,olivieri-davidson-04}, the presence of an activation sequence of 
genes can be clearly seen. We concluded that neither fully synchronous
nor completely random asynchronous network dynamics are suitable models. Thus the activation/update sequence in a RBN should be in some way related to the topology of the network, i.e. on the mutual chemical interaction structure of proteins, RNA, genes, and
other molecules which is abstracted in the network.\\
Aiming at remaining faithful to biologically plausible timing of events without introducing unnecessary complexity into the model, we considered the influence of one node on another as biological {\it activating} or {\it 
repressing} factors: only when the state of the node is turned or stays {\it on} has this node an effect on the subsequent nodes in the activation sequence. In contrast, nodes changing their state to or remaining {\it off} have no impact on 
nodes they are linked to, thus breaking the cascade. In other words, only the activation of an activator or a repressor will have a repercussion on the list of nodes to be updated at the next time-step. This update scheme, which has been briefly described previously in~\cite{darabos2007}  is called 
the Activated Cascade Update (ACU) .
\begin{figure*} [!ht]
\begin{center}
\mbox{\includegraphics[width=10cm]{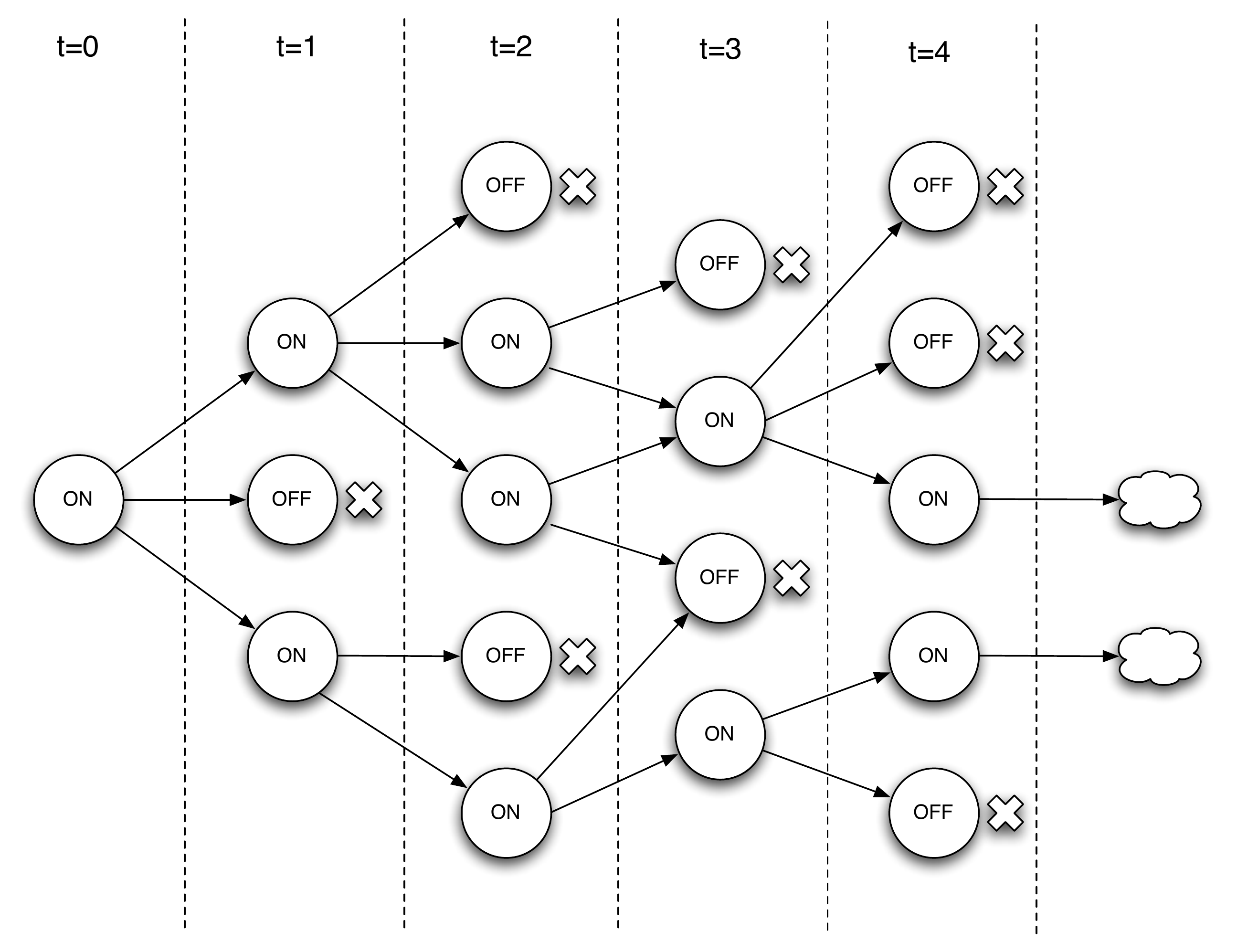} } \protect
\end{center}
\caption{A possible Activated Cascade Update sequence. At time $t=0$ a node $N_0$ is chosen at random and updated according to its inputs, if the new state of $N_0$ is inactive, another starting node $N_0$ is chosen at random. At time $t=1$ all the nodes receiving an input from $N_0$ are updated according to their own inputs, those becoming or remaining active (state {\it on}) decide which node will be updated at the next time step. The 
cascade continues according to this scheme.}
\label{acu_diagram}
\end{figure*}
As a consequence of this novel update procedure, the definition of point or cyclic attractors changes slightly, because the state of a network at any give time $t$ is, from now on, not only determined by the individual state $s_i^t \in \{ {\it on},{\it off} \} $ of each node but also by the list $l^{t+1}$ of nodes to be updated at the next time step. The concept of loose attractor has, in this context, no relevance.



\section{Methodology and Simulations }
\label{subsect:setting}

In this work we investigate the effect of the new ACU update scheme presented in Section \ref{semi-sync} vs. the previous SU on SFBNs for a set of $\gamma$ exponents of the scale-free distribution $\gamma \in \{2.0, 2.5, 
3.0\}$. The results will be 
compared to classical RBNs and all three sizes mentioned above will be studied. In order to explore their behavior in the three different regimes, we propose to vary $\bar{k} \in \{1.8, 2.0, 2.2\}$, thus keeping the probability $p$ of 
the node update functions to $p=0.5$.
In an effort to probe the network scaling properties, we have simulated ensembles of graphs with $N \in \{100, 150, 200\}$ which, although still comparatively small,  is closer to the observed GRNs sizes and
still computationally feasible. \\
For each combination of topology, update and size, we produce 50 networks. To each network, we associate 20 randomly generated sets of Boolean update functions. A network-function pair is called a realization. 
Subsequently, for each realization we create 500 different initial configurations (ICs) with equal probability for each gene to be expressed or not. Starting from each IC, we let each realization run over a number of initial steps depending on the size 
$N$ of the network (10000 for $N=100$, 20000 for $N=150$, and 30000 for $N=200$). This allows the system to possibly stabilize after a transient period, reaching the basin of an attractor. After this primary period, we determine over another 
1'000 time steps if the system has relaxed to an attractor. If so, we define the length of that attractor as the minimum number of steps necessary to cycle through the attractor's configuration. In other words, we run 50 networks 
$\times$ 20 update functions $\times 500 $ ICs $ =  5 \times 10^5$ simulations for all combination of 3 sizes, 3 regimes and 2 updates for a total of $9 \times 10^6$ simulations. Very often in this work, we will omit to show figures of all three different sizes as this parameter does not always have an impact on the results, that are in turn very similar for all sizes. Nevertheless, all sizes and cases have been thoroughly simulated and studied.


\section{Finding Attractors}
\label{subsect:analysis}

During the simulations, we have analyzed for each IC of each realization whether the system has relaxed  to a single state (point attractor) or cycled through the configurations of a periodic attractor. According to Kauffman's estimate \cite{kauffman93}, the median lengths of attractors $l \propto \sqrt{N}$ or linear at most for $\bar{k}$ critical. For $\bar{k}$ well into the chaotic regime, the median length grows exponentially with $N$. Biologically speaking, very long attractors are unlikely to have any meaning due to the actual gene expression time which is in the order of seconds to minutes. Therefore we investigate in depth only attractors with lengths ranging from 1 to 100 states. 
Admittedly, the maximum length is arbitrary but remember that, according to Kauffmann, we are mostly interested in attractors that are short and stable in the ``critical'' regime (or ``edge of chaos''). In natural systems, point and periodic attractors may have different significations. As an example periodic attractors can be interpreted as a model of the genetic regulatory system during the cell cycle, whereas point attractors can refer to the end of the differentiation cycle of a stem cell. Although it has been shown that point attractors may play a fundamental role outside the stem cell context, as in the works of Albert {\it et al.}~\cite{albert-2003-223} and more recently of \'Alvarez-Buylla and coauthors \cite{alvarezbuylla2008}, we will often present simulation results and statistics both with and without point attractors. The reason for this is that in some instances under ACU, the scheer number of point attractors tend to bias the statistics and to make the results more difficult to interpret (see Fig. \ref{attractorsFreq}). 

\subsection{Number of Attractors}
\label{nbAtt}

In Fig. \ref{nb_attractors} we show the frequencies at which networks of size (a) $N=100$ and (b) $N=200$ find attractors of any length. Since the simulations for networks with $N=150$ nodes behave similarly to larger and smaller ones, we do not show them here. 
\begin{figure*} [!ht]
\begin{center}
\begin{tabular}{cc}
\mbox{\includegraphics[width=6.5cm]{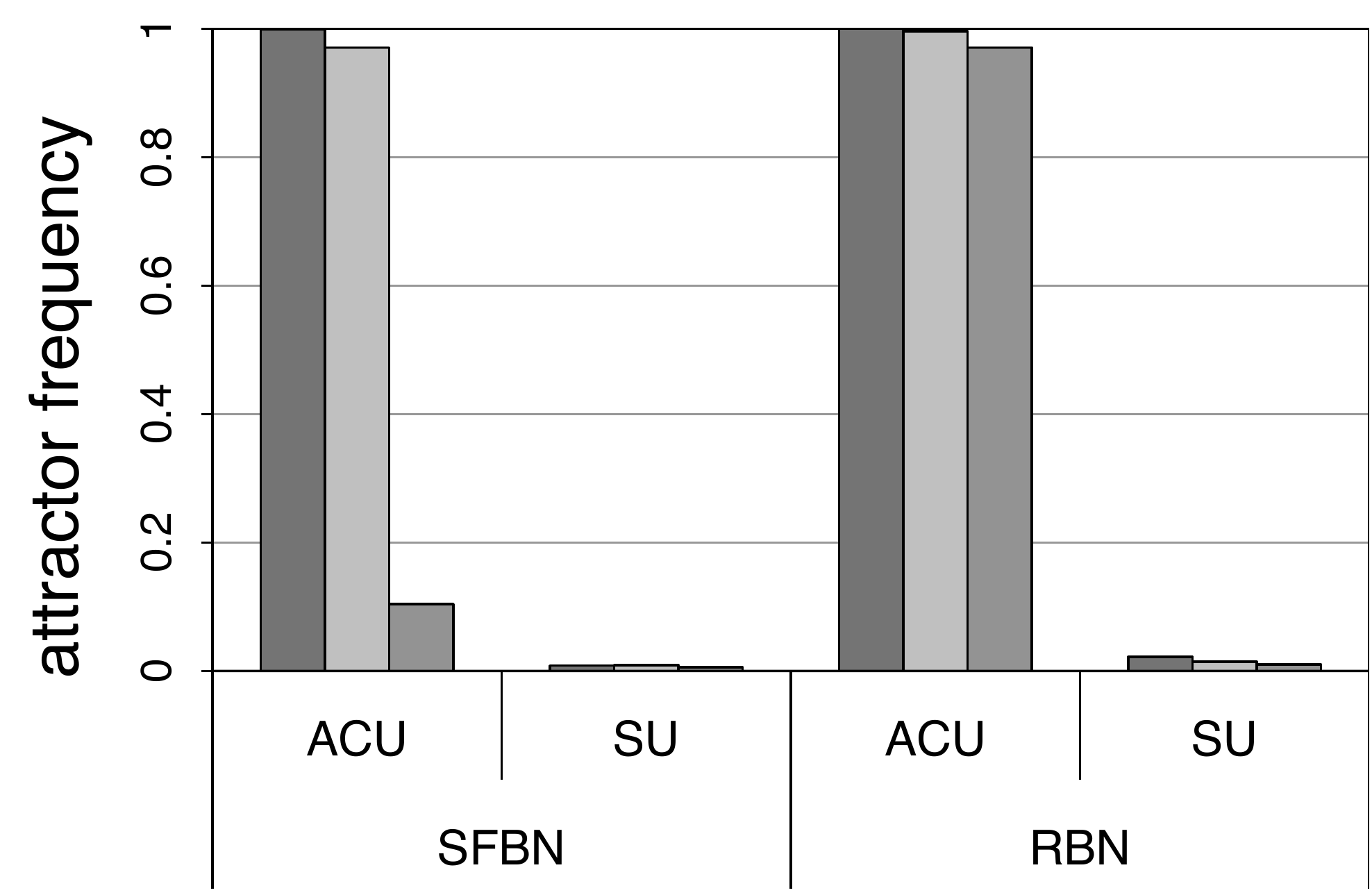} } \protect &
\mbox{\includegraphics[width=6.5cm]{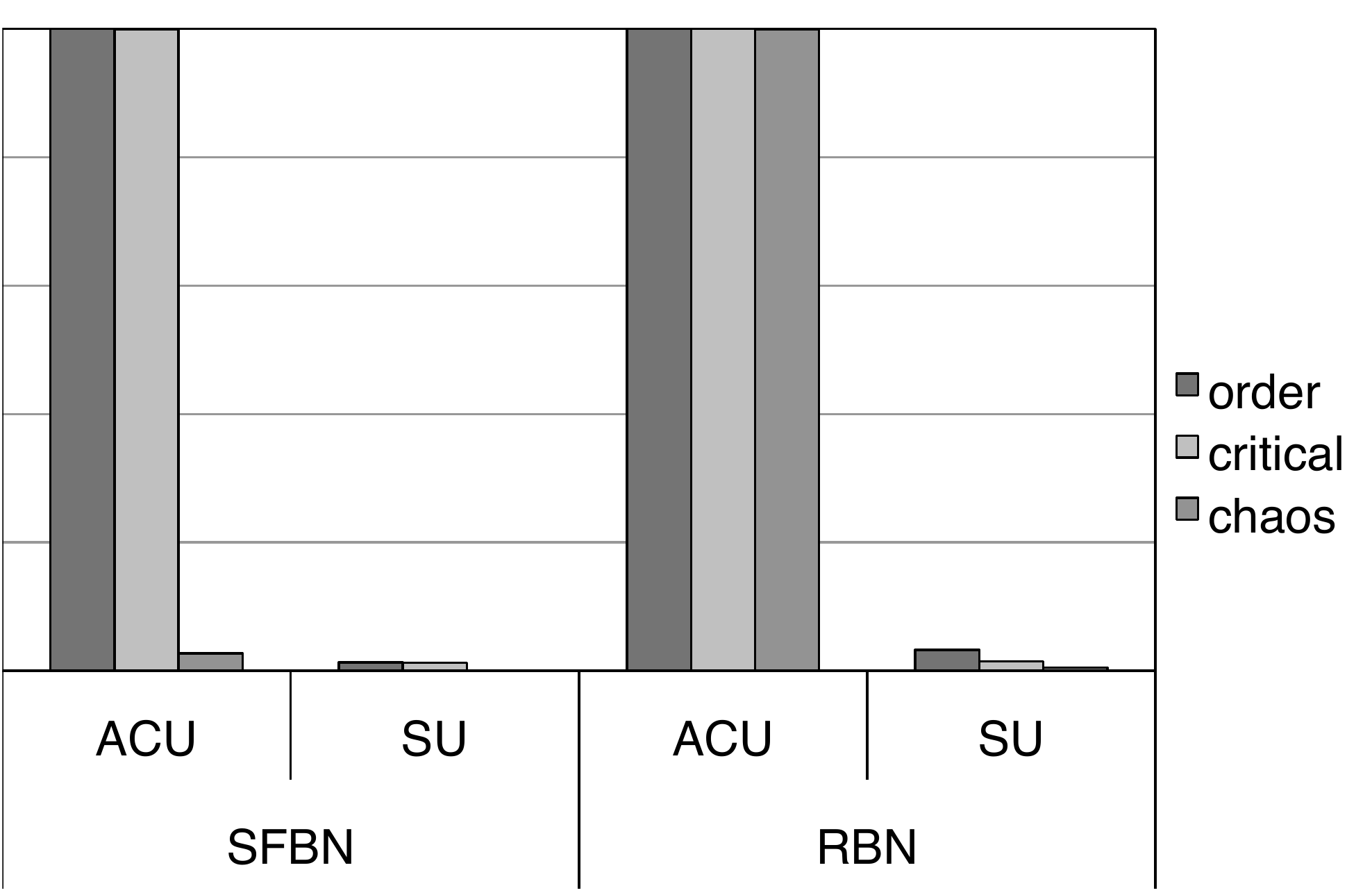} } \protect \\
(a)  &  (b) \\
\end{tabular}
\end{center}
\caption{Number of attractors of any length. Comparing the frequency at which simulations have found an attractor of any length for realizations with (a) $N=100$ and (b) $N=200$ genes. We compare all network topologies and update schemes.}
\label{nb_attractors}
\end{figure*}
Fig. \ref{nb_attractors} shows that almost all instances under ACU we find an attractor, except for scale-free systems in a chaotic regime,
which tend to produce $10$ to $50$ times less attractors. On the contrary, GRNs under SU struggle to relax to an attractor. In both RBNs and GBNs, we observe that the number of attractors does not seem to be impacted by the scaling. \\
Frequencies and length concerning simulations of shorter and more biologically plausible attractors are shown below in Fig. \ref{attractorsFreq} and in Fig. \ref{attractorsLength} respectively. On the right-hand sides, point attractors have been removed from statistics.
\begin{figure*} [!ht]
\begin{center}
\begin{tabular}{ccc}
& $1 \le l_{att} \le 100$ & $2 \le l_{att} \le 100$\\

\raisebox{10ex}[0pt]{\begin{sideways}$N=100$\end{sideways}} 
&\mbox{\includegraphics[width=6cm]{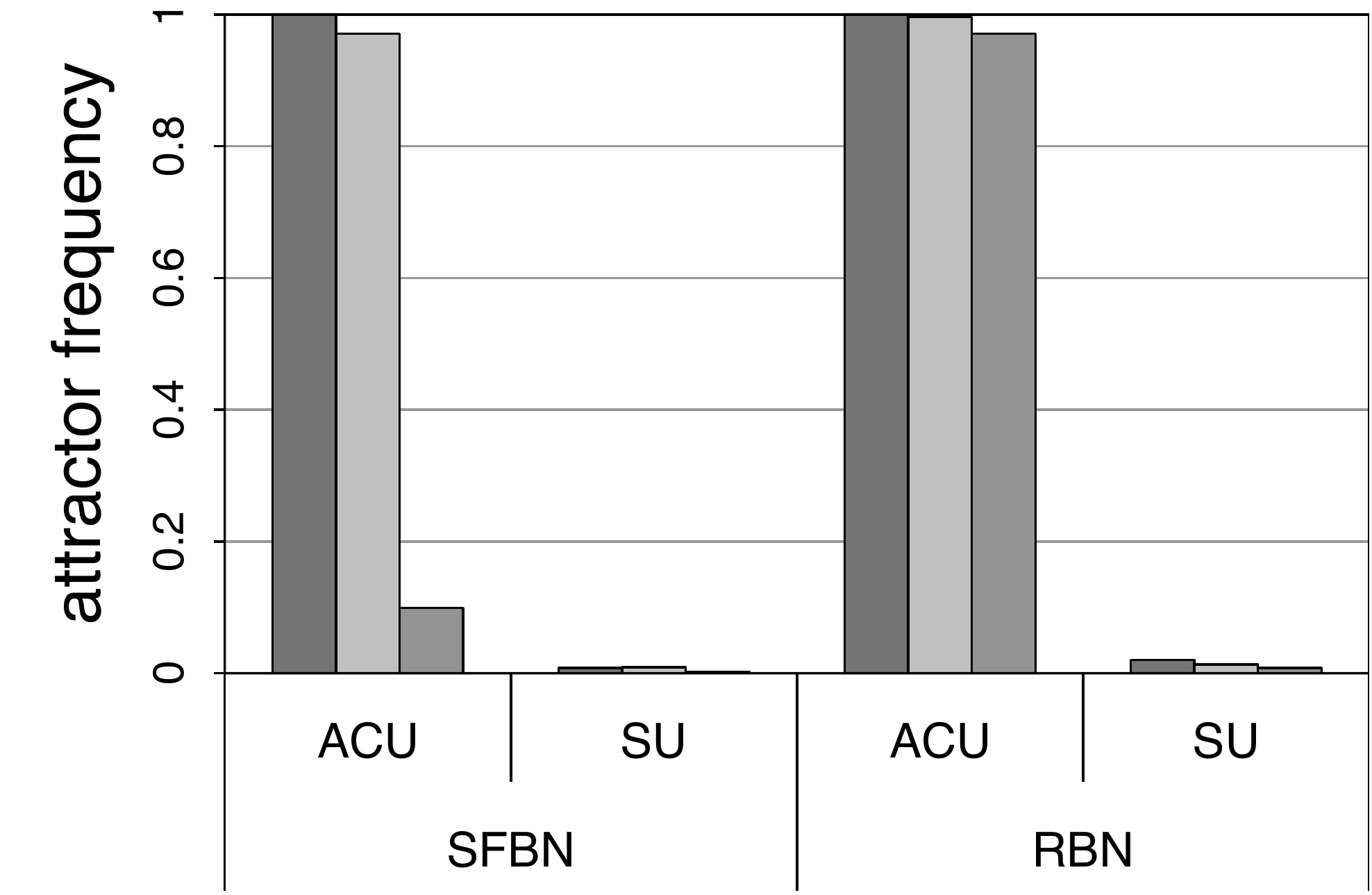} } \protect 
&
\mbox{\includegraphics[width=6cm]{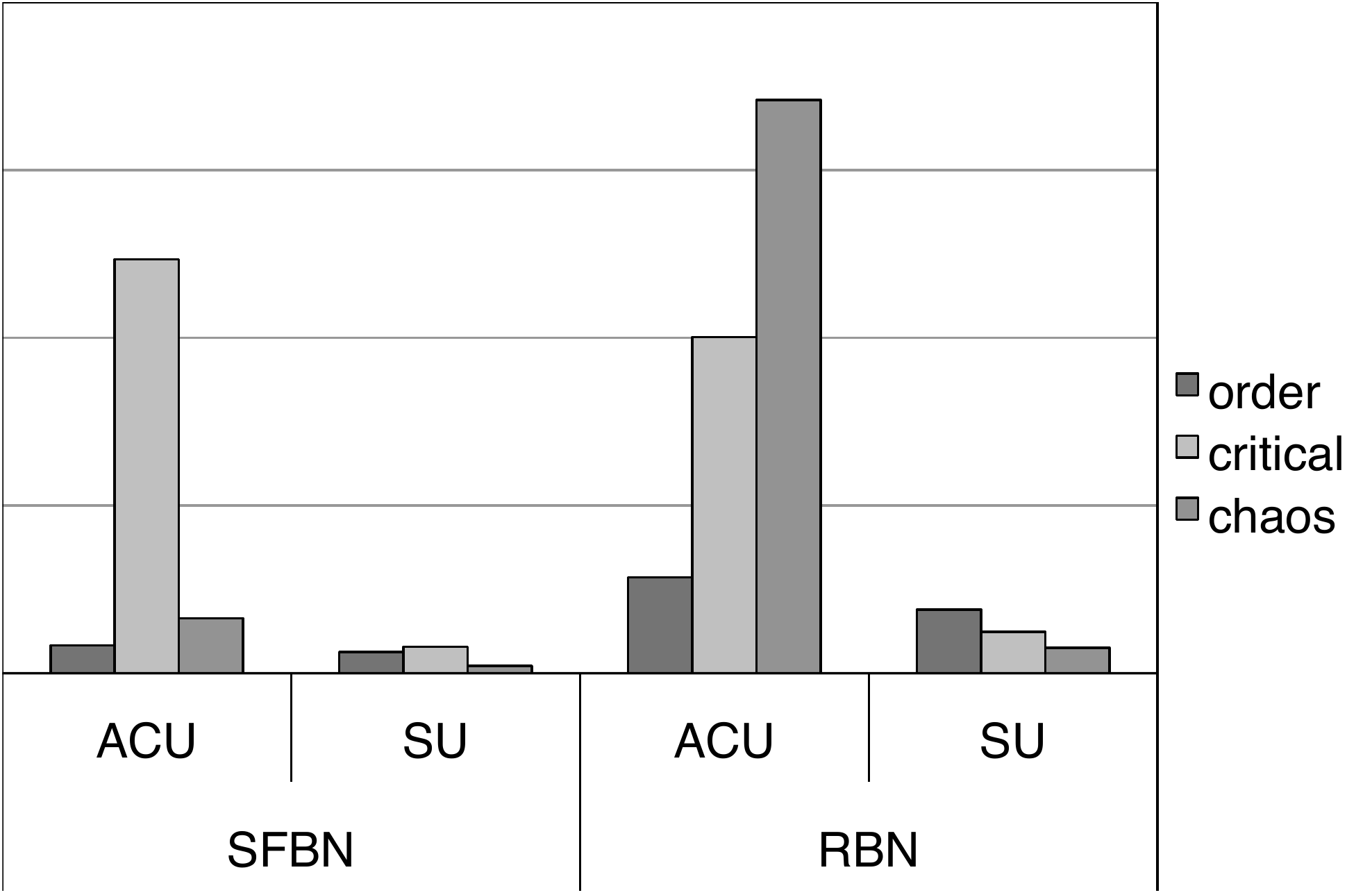} } \protect \\
&(a)  &  (b) \\

\raisebox{10ex}[0pt]{\begin{sideways}$N=200$\end{sideways}} 
&\mbox{\includegraphics[width=6cm]{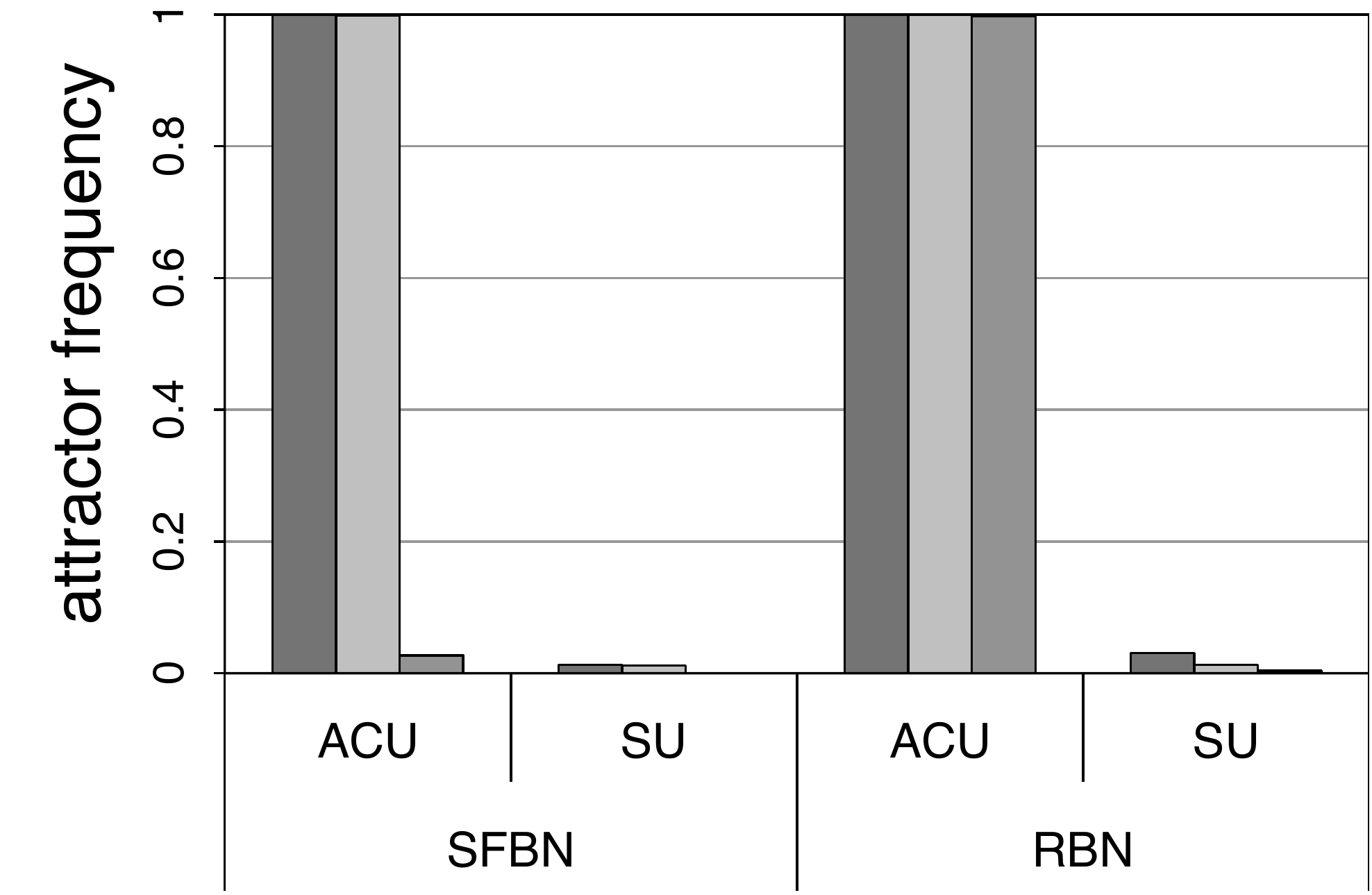} } \protect 
&
\mbox{\includegraphics[width=6cm]{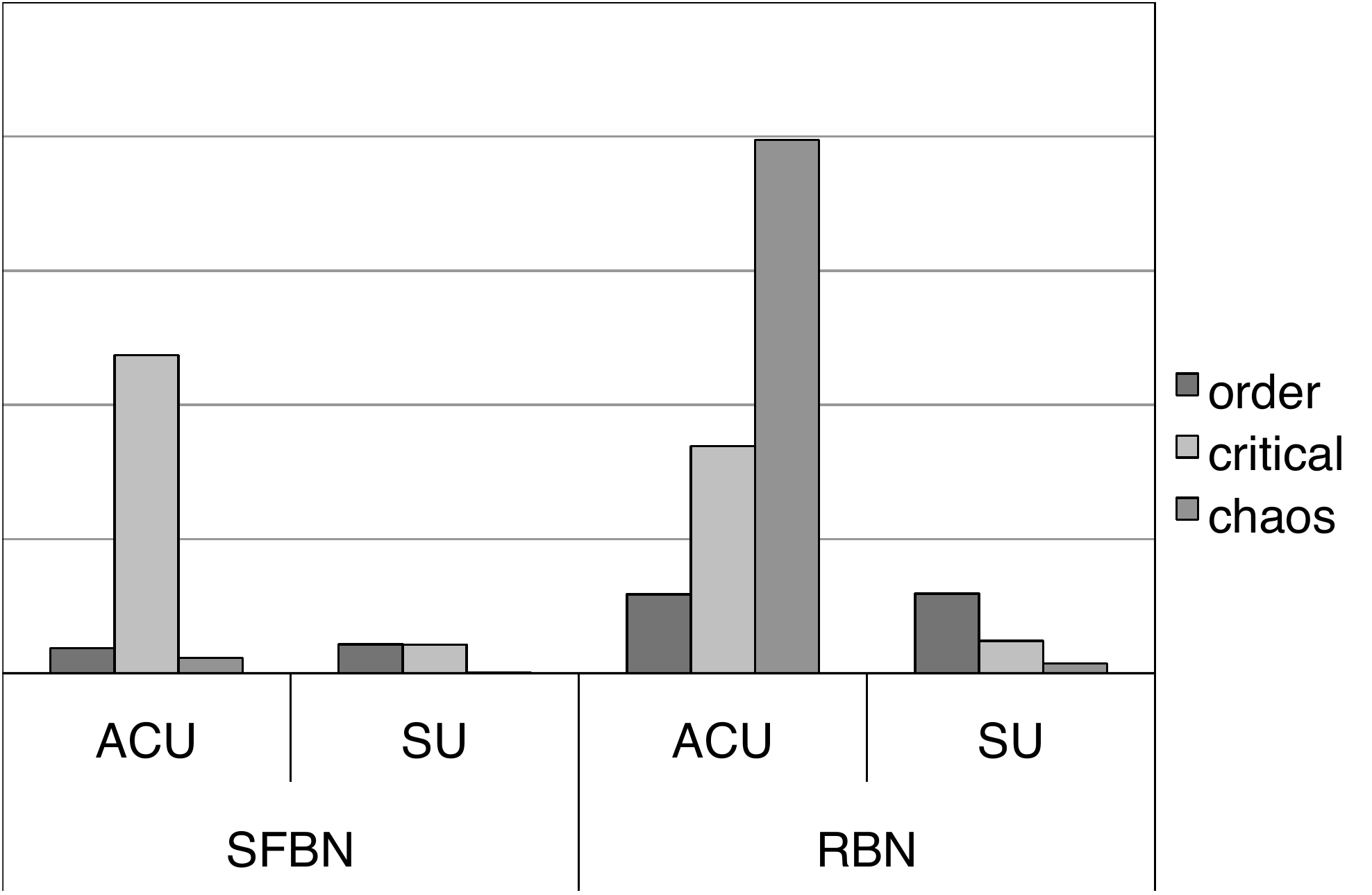} } \protect \\
&(c)  &  (d) \\
\end{tabular}
\end{center}
\caption{Number of attractors of biologically plausible lengths. Comparing the frequency at which simulations have found attractors of length (a)(c) between 1 and 100 and (b)(d) between 2 and 100. We consider networks of size (a)(b) $N=100$ and (c)(d) $N=200$. }
\label{attractorsFreq}
\end{figure*}
When comparing Figs. \ref{nb_attractors}(a) with \ref{attractorsFreq}(a) and \ref{nb_attractors}(b) with \ref{attractorsFreq}(c) respectively, there is virtually no difference, as over 95\% of the attractors are in fact below a length of 100 states. As for attractors of length between 2 and 100 in Fig. \ref{attractorsFreq}(b) and (d), we see that ACU systems, whether scale-free or random, find more attractors than those under SU. We note that SFBNs in a critical regime under ACU have a peak in finding attractors, compared with other regimes, which are exactly the attractors we are interested in. In RBNs and GBNs, we observe that the number of attractors does not seem to be impacted by the scaling either. Using ACU almost every IC of every realization leads to an attractor, no matter what the regime is. On the contrary, under SU the overall number of attractors tends to decrease as the system goes from order to chaos.

\subsection{Variety of the Attractors}
\label{variety}

Table \ref{attUsage} shows how many times on average the same attractor has been found for each update scheme, regime and topology over the 500 ICs the system has been submitted to. We divided the results for attractors including and excluding point attractors (PA).
\begin{table}
\begin{center}
\begin{tabular}{|rrr|cc|cc|}
\hline
&&&			\multicolumn{2}{c}{N=100} & \multicolumn{2}{c}{N=200} \vline \\
&&&			w PA  &	w/o PA & w PA  &	w/o PA\\
\hline
SFBN	&ACU	&order	&1.01	&1 		& 1		& 1\\
		&		&critical	&1.03	&1.13	& 1.01	& 1.01\\
		&		&chaos	&4.42	&10.44 	& 2.16 	& 1.68\\
		&SU		&order	&121.16	&73.37	& 46.55 	& 78.77\\
		&		&critical	&104.26	&95.82	& 60.54	& 65.46\\
		&		&chaos	&5.08	&43.60	& 1		& 1\\
\hline
RBN		&ACU	&order	&1.01	&1		&1		& 1\\
		&		&critical	&1.01	&1.01	&1		& 1\\
		&		&chaos	&1.03	&1.07	&1.01	& 1.01\\
		&SU		&order	&44.90	&39.39	&22.24	& 26.33\\
		&		&critical	&58.01	&53.26	&30.34	& 33.93\\
		&		&chaos	&61.01	&58.75	&24.69	& 30.04\\
\hline
\end{tabular}
\end{center}
\caption{Attractors diversity. The average time each attractor has been found over 500 ICs. Cases with (w PA) and without (w/o PA) point attractors are segregated. In this case, $N=100$ and $200$, and the attractors length is limited to a maximum of 100 states.}
\label{attUsage}
\end{table}
We can summarize in Table \ref{attUsage} a few observations as follows: the topology type does not seem to have a drastic effect on how often the system relaxes to the same attractor. On the other hand, the update scheme affects the total number of times the same attractor is found, and so does the regime, but in a much milder manner. In fact, we can see that SU tends to find much more often the same attractor than ACU does. In addition, we see in Figures \ref{nb_attractors} and \ref{attractorsFreq} that this SU also tends to find many fewer attractors overall. Alternatively, systems under ACU find a greater number of different attractors, only in the chaotic regime, where the overall number of distinct attractors is already very small compared with the other ones. We witness an increase in the average number of the times the same attractor is found. Note that in the chaotic regime for systems where $N=200$ under SU, the low repetition value is due to the fact that very few attractors are found. 

\subsection{Length of the Attractors}
\label{attLen}

Fig. \ref{attractorsLength} shows statistics on the length of attractors. We exclude point attractors for figures on the right-hand side (figures (b) and (d)). The bar at the center of the box is the median of the attractors lengths, the upper and lower box delimiters are the third and first quartile respectively. The whiskers show extreme minimal and maximal values. Results are shown only for the case where the networks size $N=200$, as results for smaller sizes are very similar.
\begin{figure*} [!ht]
\begin{center}
\begin{tabular}{ccc}
& $1 \le l_{att} \le 100$ & $2 \le l_{att} \le 100$\\

\raisebox{10ex}[0pt]{\begin{sideways}SFBN\end{sideways}} 
&\mbox{\includegraphics[width=6cm]{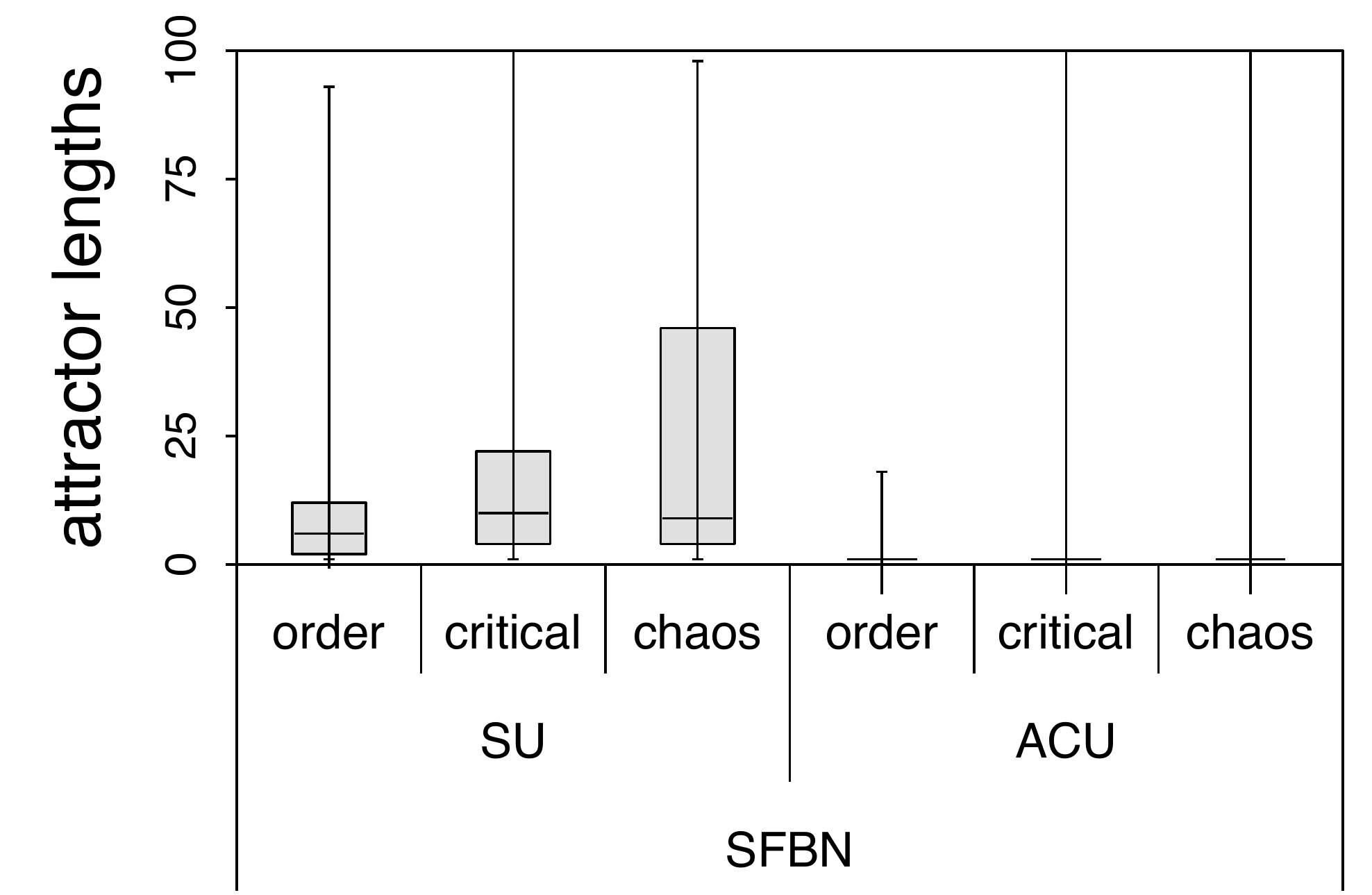} } \protect 
&
\mbox{\includegraphics[width=6cm]{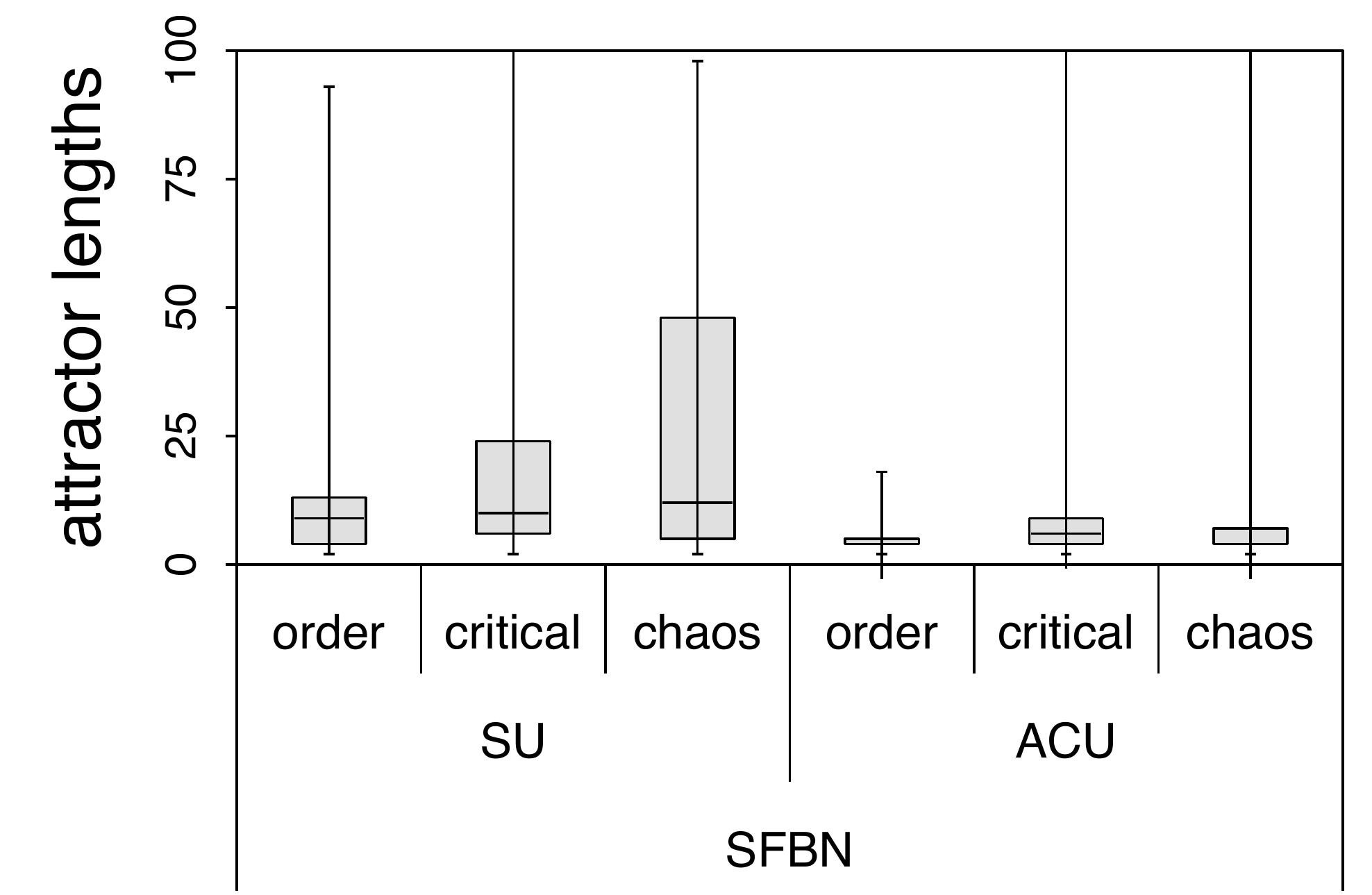} } \protect \\
&(a)  &  (b) \\

\raisebox{10ex}[0pt]{\begin{sideways}RBN\end{sideways}} 
&\mbox{\includegraphics[width=6cm]{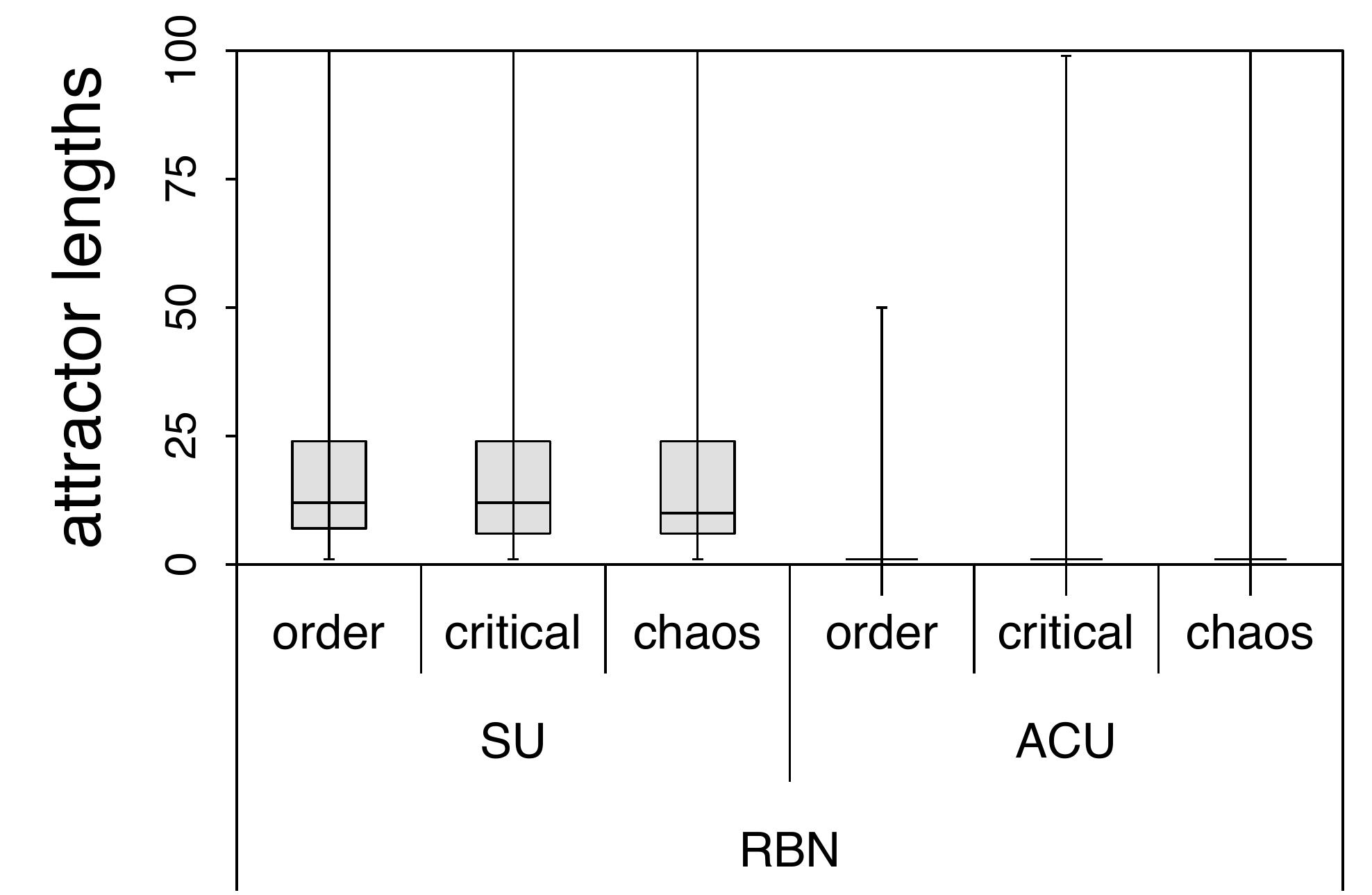} } \protect 
&
\mbox{\includegraphics[width=6cm]{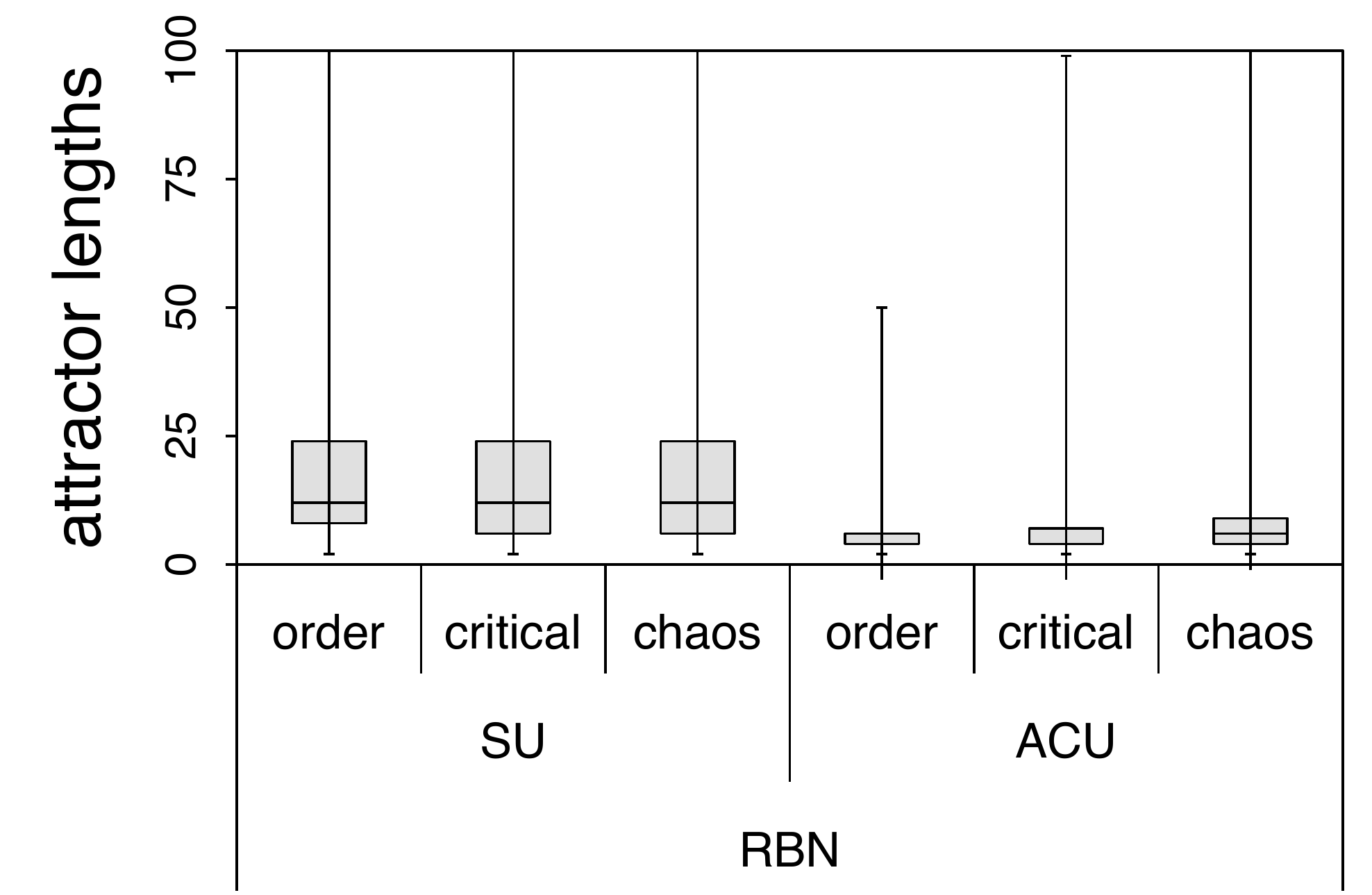} } \protect \\
&(c)  &  (d) \\
\end{tabular}
\end{center}
\caption{Attractors lengths. Comparing the length of the attractors found by simulations. In the left-hand side column (a)(c) we show statistics of attractors of size up to $100$ states and the right hand-one (b)(d) also excludes point attractors. Horizontally, the upper row (a)(b) corresponds to SFBNs and the lower row (c)(d) to RBNs. All systems have $N=200$ nodes. }
\label{attractorsLength}
\end{figure*}
Once more we see that scaling does not have a significant impact on the length of the attractors that are found by the systems. It is mostly the regime the system evolves in and, in a lesser manner, the update scheme that 
defines the attractors average length. We note in Fig. \ref{attractorsLength}(a) and (c) that, although under-represented, attractors under SU seem to be the longest, especially in the chaotic regime. When focusing on the more interesting part of the attractors population in Fig. \ref{attractorsLength}(b) and (d), we see that the lengths remain comparable, though slightly shorter when considering systems under ACU. We also know from the section \ref{nbAtt} above that those attractors are much more frequent in systems under ACU. A global conclusion concerning the attractors distribution is that the update model has a prominent effect on the number and length of attractors over the networks topologies.\\
Fig. \ref{attLenDistr} shows the distribution of the number of attractors according to their length on a \emph{log-log} scale.
\begin{figure*} [!ht]
\begin{center}
\begin{tabular}{ccc}
& ACU & SU\\

\raisebox{10ex}[0pt]{\begin{sideways}SFBN\end{sideways}} 
&\mbox{\includegraphics[width=6cm]{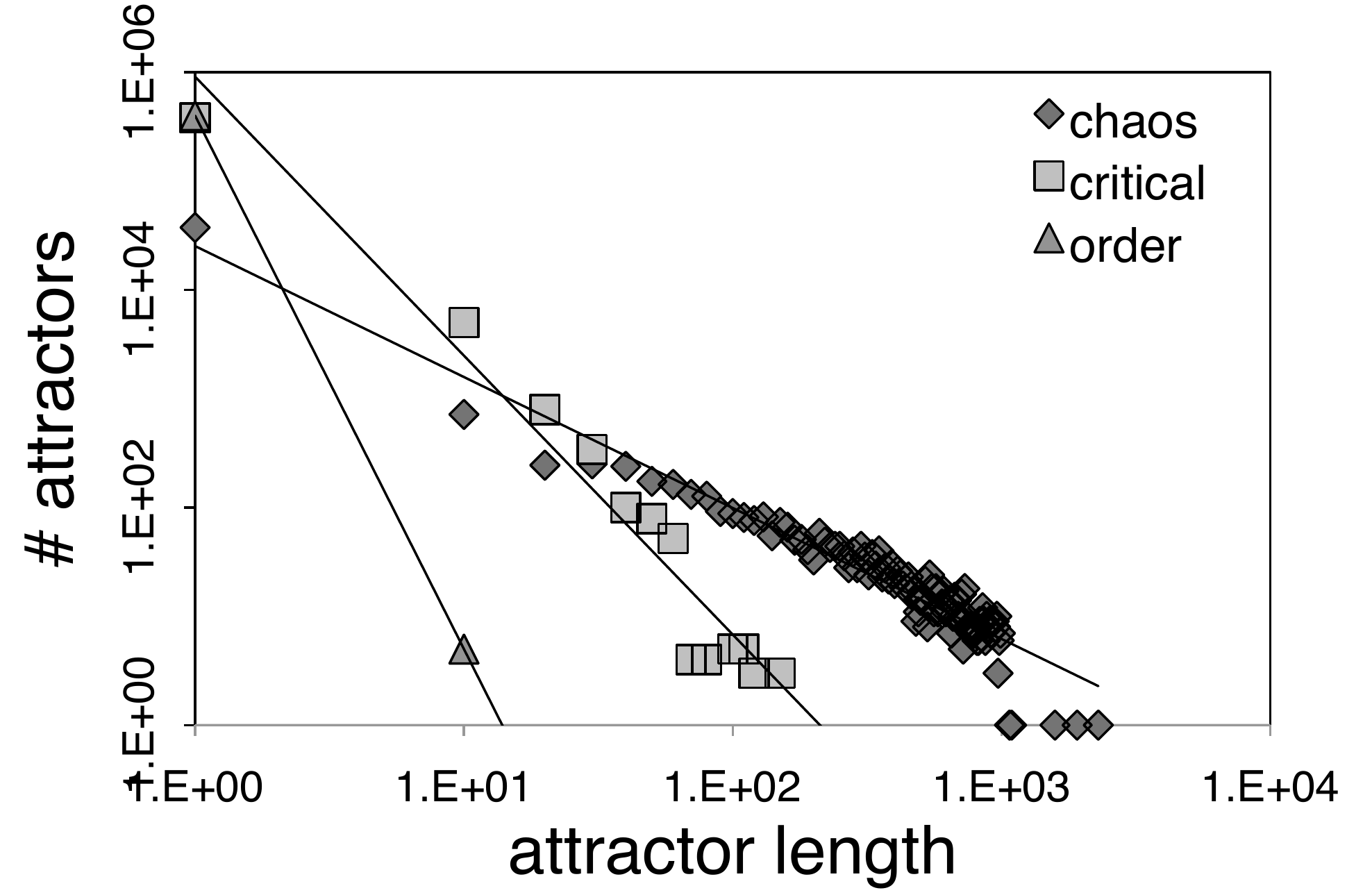} } \protect 
&
\mbox{\includegraphics[width=6cm]{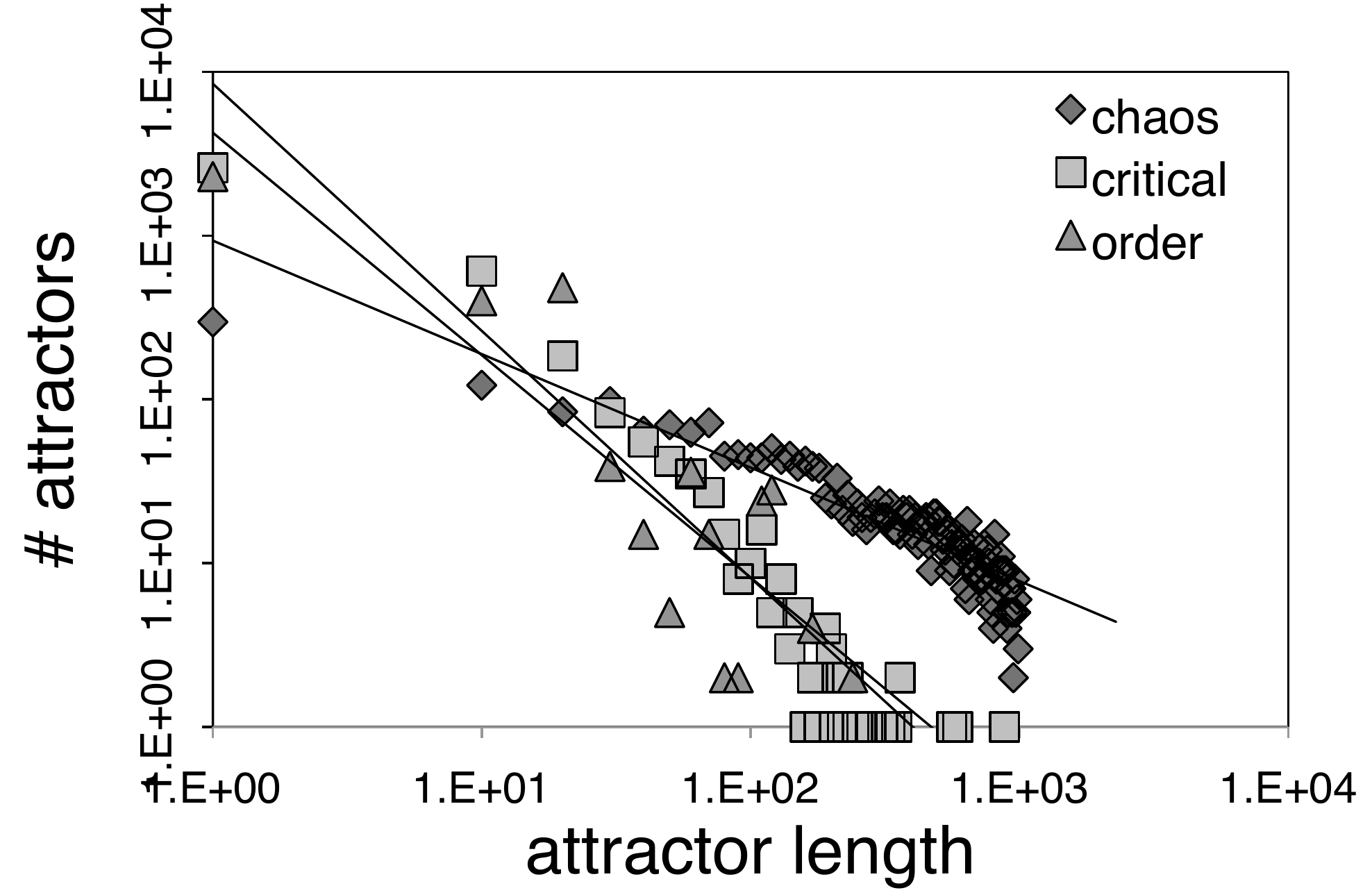} } \protect \\
&(a)  &  (b) \\

\raisebox{10ex}[0pt]{\begin{sideways}RBN\end{sideways}} 
&\mbox{\includegraphics[width=6cm]{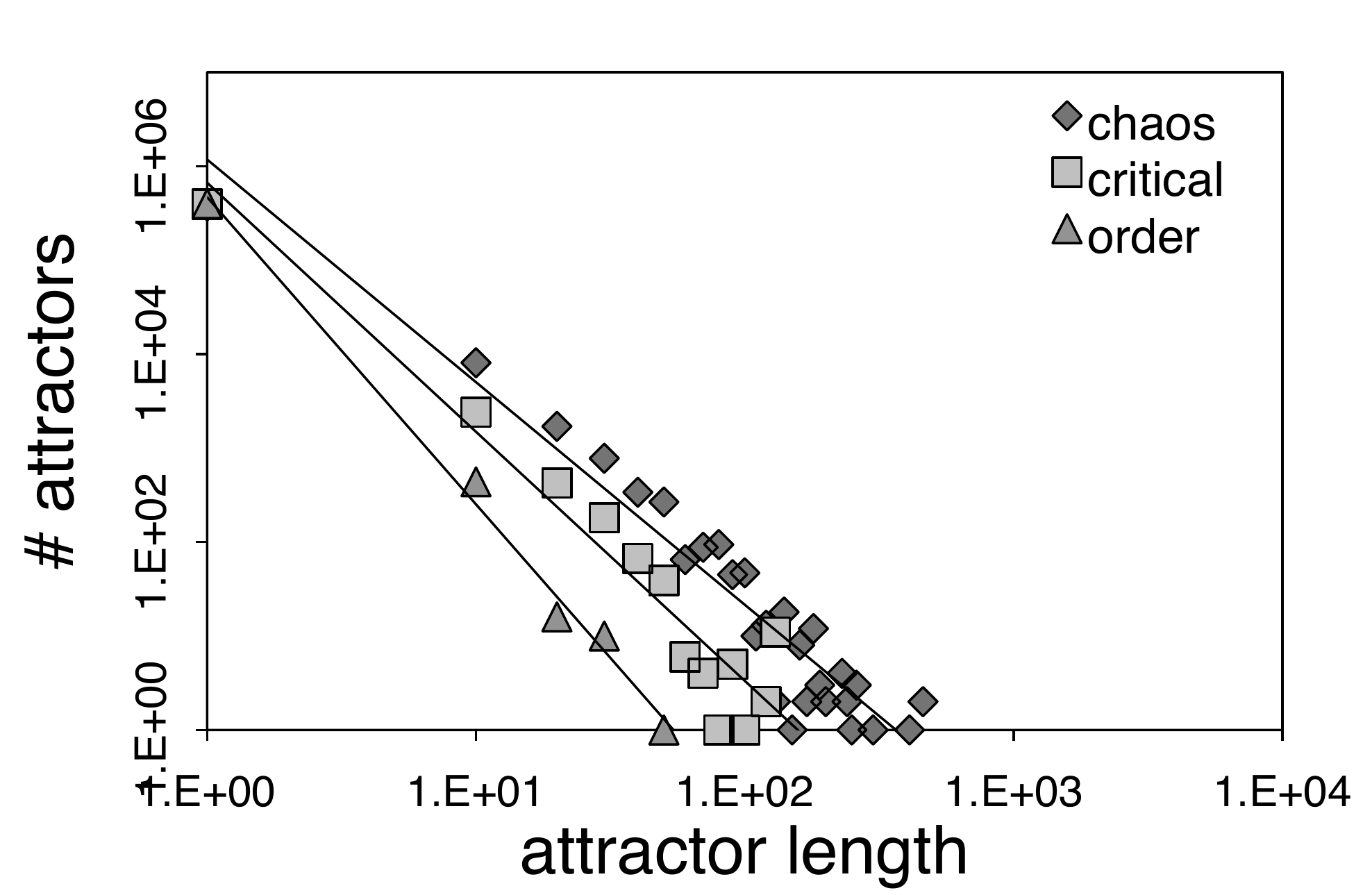} } \protect 
&
\mbox{\includegraphics[width=6cm]{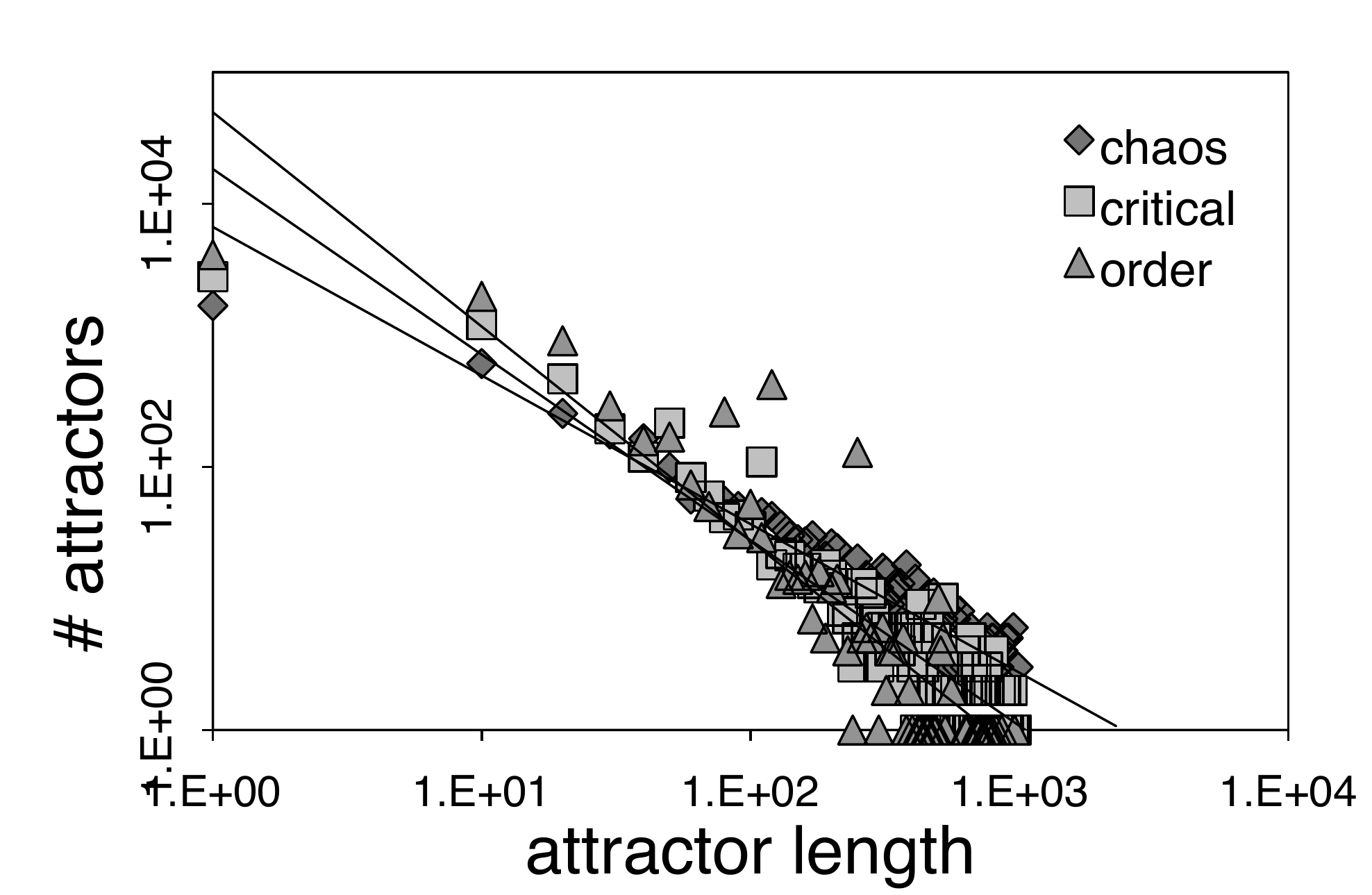} } \protect \\
&(c)  &  (d) \\
\end{tabular}
\end{center}
\caption{Attractors lengths distribution. Distribution of the attractors' lengths between 1 and 100 states. The left-hand side column represents systems under (a)(c) Activated Cascade Update and the right-hand side on (b)(d) Synchronous Update. Figures in the upper row (a)(c) show results for SFBN, the lower one (b)(d) for RBN networks of size N=100. Note that the vertical axis in the four figures have different scales. The continuous lines are power regressions to be used as a guide for the eye.}
\label{attLenDistr}
\end{figure*}
Interestingly, the distribution of attractors lengths for SFBNs shows a \emph{long-tail} for both updates, which is especially marked for systems in the chaotic regime. But this distribution does show comparable tendencies on SFBNs under any update, whereas for RBNs, the tail under ACU is much less pronounced than it is under SU. So we see that now, regime has a greater influence on attractors length for SFBNs and not the update scheme. It is the opposite for RBNs, where the timing of update has a greater impact. All cases compare favorably with Aldana's work \cite{aldana-sf-03} where he clearly shows that SFBN systems, although he had the input and the output distribution swapped, exhibit a long tailed distribution of the attractors lengths in the chaotic regime. This phenomenon is much less pronounced in the ordered and critical regime. Although we observe a difference in the case of RBNs, the different regimes are difficult to tell apart. Nevertheless, the distributions also show a power-law-like curve for all regimes, with a tail longer than that of SFBNs in the ordered and critical regime. This second observation is in line with Iguchi et al.~\cite{iguchi07} where, in the case of smaller RBNs, both in the critical and chaotic regime, attractors lengths distributions show a long tail. We also notice that ACU has the unexpected effect to help tell apart regimes, as distribution is much easier to distinguish than under SU.


\subsection{Scaling}
\label{sect:scaling}

Modern high throughput technologies for genetic analysis have tremendously contributed to the unveiling of ever bigger parts of GRNs in living organisms. Present sub-networks sizes range from a few tens to a few hundreds of genes. In the section above, we thoroughly investigated the attractor's dynamics of systems of sizes ranging from 100 to 200 nodes, and have noticed the size of the system mainly affects the number of attractors that are found. This fact was expected as the state space grows with the number of nodes $N$ as $2^N$, making it harder for the system to relax in a cycle. For other properties such as the length distribution or mean length, although sightly different, the general tendencies are not impacted by the scaling. \\
In order to study the effect of scaling on Boolean systems and its effect on both different topologies and both updates, we have extended the above analysis to networks of size $N=25, 50, 75, 100, 125, 150, 175, 200$. Due to extreme computational resources necessary, we have unfortunately not been able to increase $N$ to greater sizes and obtain sufficiently reliable data. Indeed, as the number of node grows, the transient period before the system reaches an attractor and length of the attractors themselves increases dramatically, especially in the chaotic phase. Aldana \cite{aldana-sf-03} shows the increase in the transient time and also shows trends on the expected length of the attractors as $N$ grows.\\
Fig. \ref{figScaling} shows the trends followed by the attractors lengths as the size of the systems grows for both topologies, updates and all three regimes.
\begin{figure*} [!ht]
\begin{center}
\begin{tabular}{ccc}
& ACU & SU\\

\raisebox{10ex}[0pt]{\begin{sideways}SFBN\end{sideways}} 
&\mbox{\includegraphics[width=6cm]{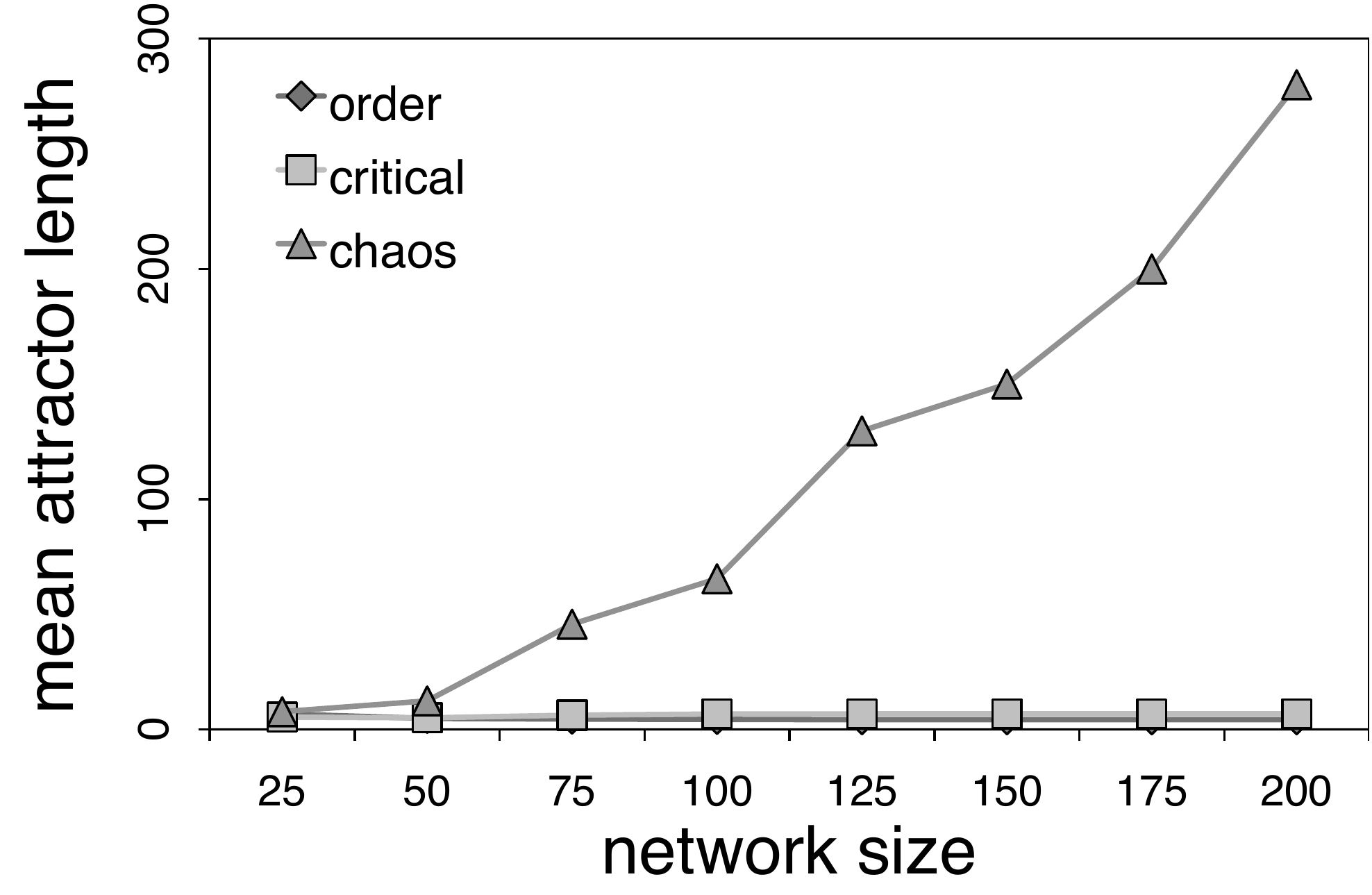} } \protect 
&
\mbox{\includegraphics[width=6cm]{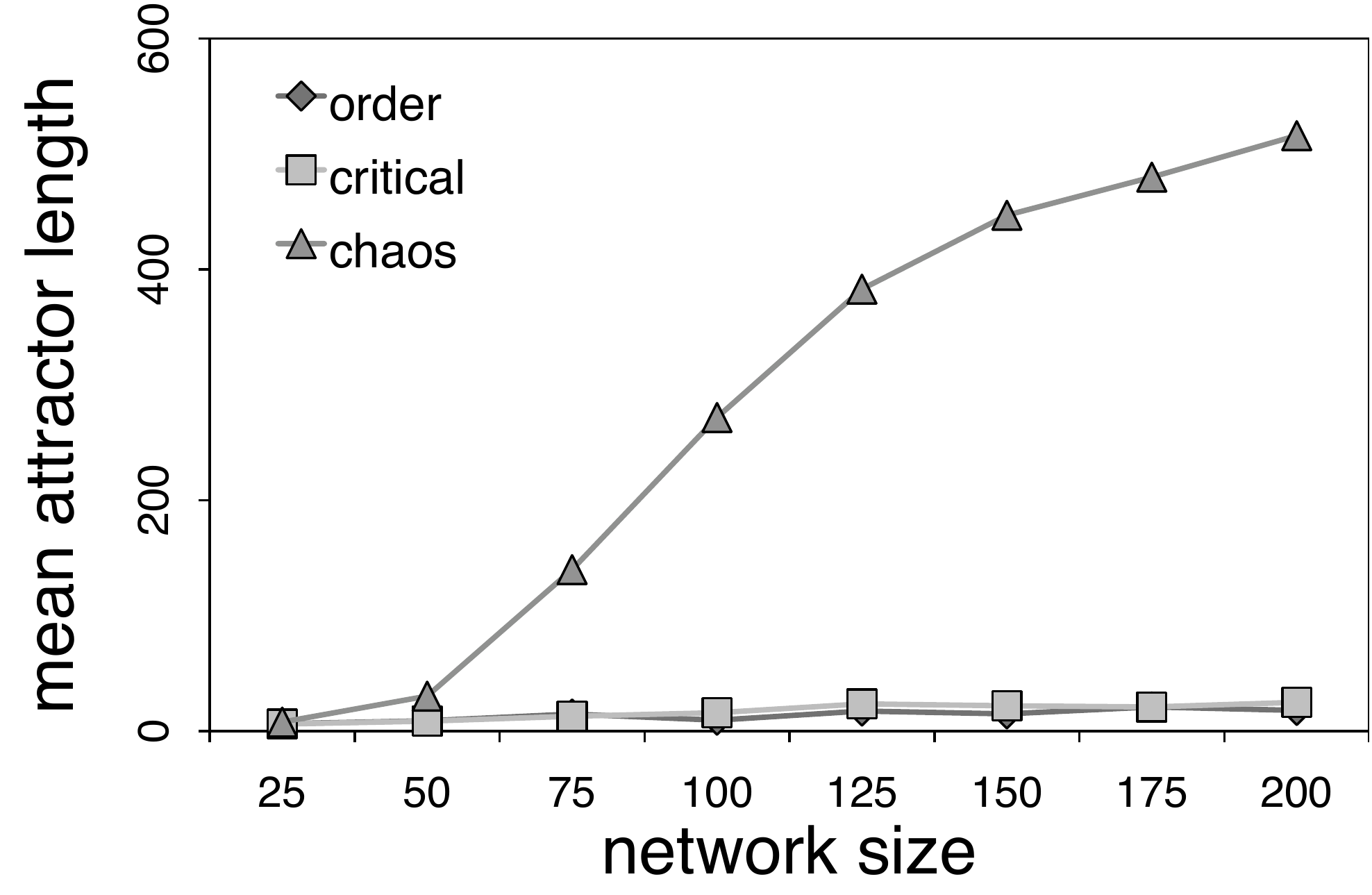} } \protect \\
&(a)  &  (b) \\

\raisebox{10ex}[0pt]{\begin{sideways}RBN\end{sideways}} 
&\mbox{\includegraphics[width=6cm]{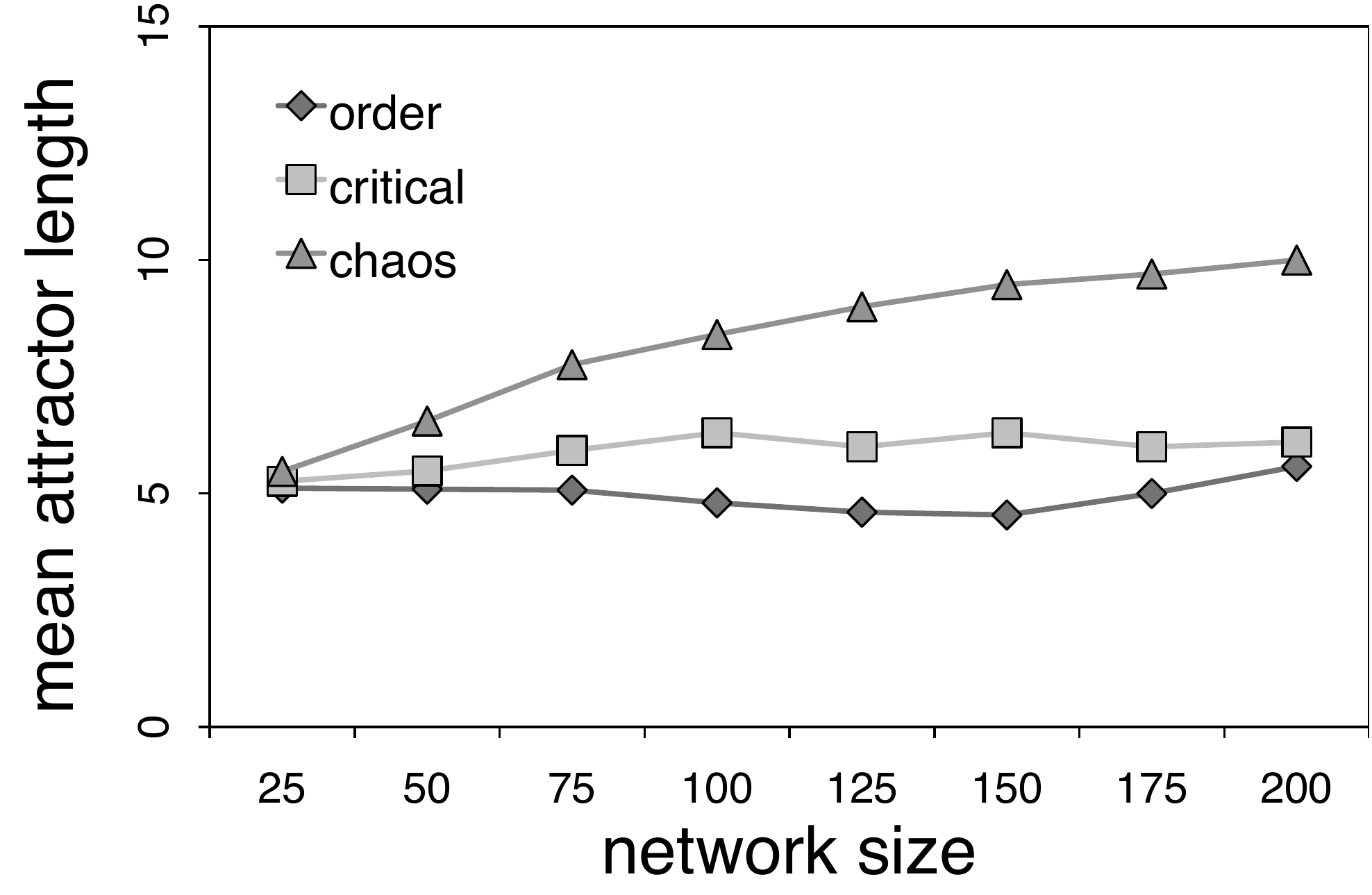} } \protect 
&
\mbox{\includegraphics[width=6cm]{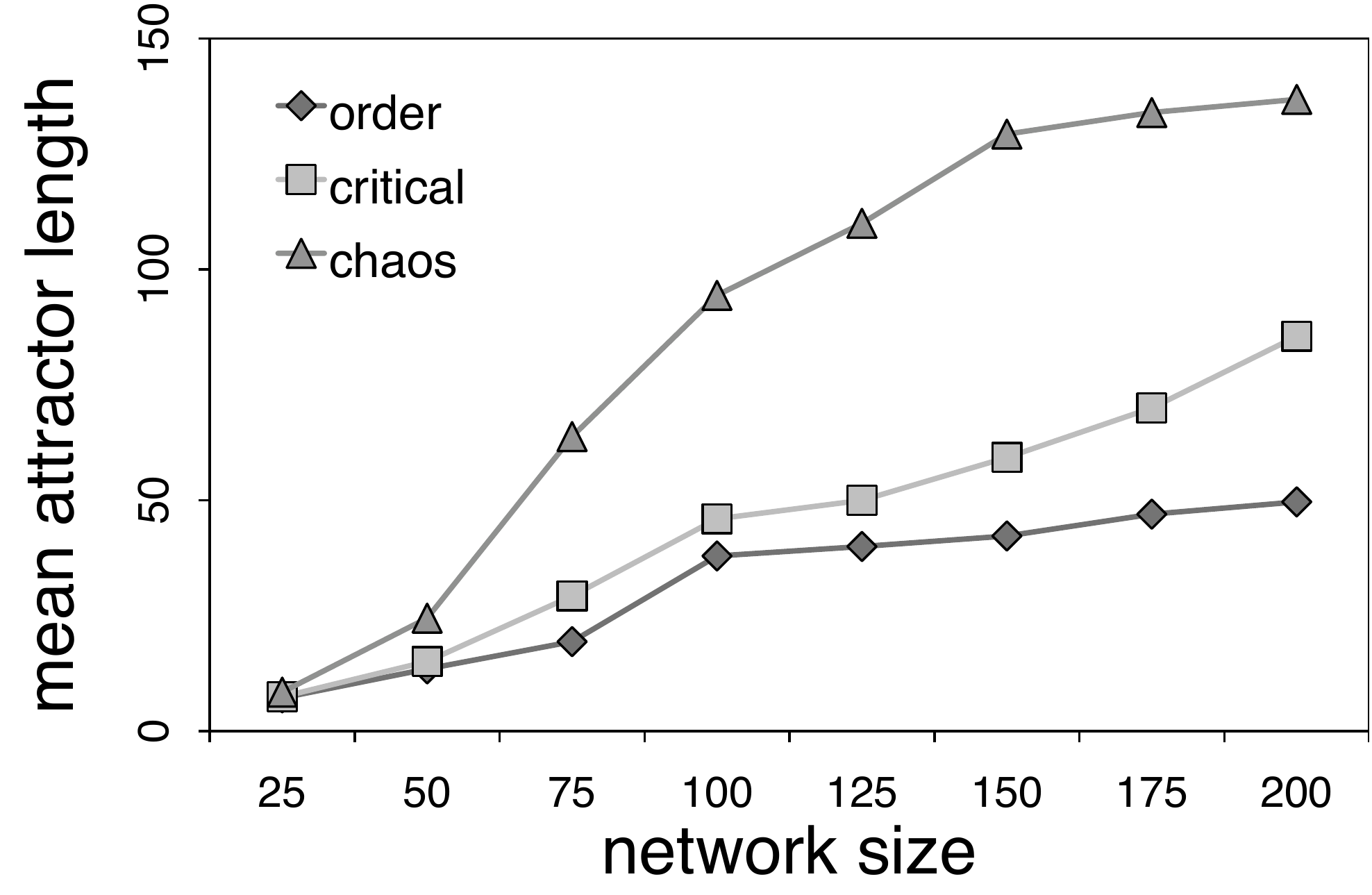} } \protect \\
&(c)  &  (d) \\
\end{tabular}
\end{center}
\caption{Attractors average cycle length with respect to the network size. Scaling of the average length of attractors compared to the networks size $N$ for (a)(b) SFBNs and (c)(d) RBNs under (a)(c) ACU and (b)(d) SU. Note that the vertical axis in the four figures have different scales. Continuous lines are only added as a guide for the eye.}
\label{figScaling}
\end{figure*}
The size of the attractors for SFBN systems under both update strategies, scales as expected form Aldana's work \cite{aldana-sf-03}. He exhaustively studied SFBNs under SU of sizes $N \in \{8,\ldots,20\}$. We witness, both for SU and ACU, a similar and expected trend, where only the length of attractors found by systems in the chaotic phase increase significantly with $N$. The mean attractor length of systems under ACU is much shorter than that of systems under SU, which is in line with the thorough analysis conducted in Section \ref{attLen}. In the case of classical RBNs, the differences in size between the regimes, although existing, is much less pronounced. The range average size is yet again much greater with SU. Under both updates, lengths for ordered and critical regimes remain relatively close whereas for the chaotic regime, the difference with the other regimes augments significantly. We have unfortunately not been able to compare the number of attractors to Aldana's work because we are only sampling much bigger systems, up to ten times larger, that cannot be exhaustively analysed in a reasonable amount of time. Nevertheless, this comforts us in the idea that our model, while in our eyes is more realistic, still it has behaviors that are in accordance with our predecessors validated work. Iguchi et al.~\cite{iguchi07} have conducted similar experiment on a limited sample of scale-free input and output distribution networks of very lage size under SU. Though their results seem in agreement with our own findings, their model is too different to draw direct parallels. 


\section{Fault Tolerance of Random Boolean Networks}
\label{sect:fault}

Failures in systems can occur in various ways, and the probability of some kind of error increases dramatically with the complexity of the systems. They can range from a one-time wrong output to a complete breakdown and 
can be system-related or due to external factors. Living organisms are robust to a great variety of genetic changes, and since RBNs are simple models of the dynamics of biological interactions, it is interesting and legitimate to ask questions about their fault tolerance aspects. \\
Kauffman \cite{kauffman2000} defines  one type of perturbation to RBNs as ``gene damage'', that is the transient reversal of a single gene in the network. These temporary changes in the expression of a gene are extremely 
common in the normal development of an organism. The effect of a single hormone can transiently modify the activity of a gene, resulting in a growing cascade of alternations in the expression of genes influencing each 
other. This is believed to be at the origin of the cell differentiation process and guides the development.\\
The effect of a gene damage can be measured by the size of the avalanche resulting from that single gene changing its behavior from active to inactive or vice-versa. The size of an avalanche is defined as the number of 
genes that have changed their own behavior at least once after the perturbation happened. Naturally, this change of behavior is compared to an unperturbed version of the system that would be running in parallel. The size of 
the avalanche is directly related to the regime in which the RBN is; in the ordered regime, the cascades tend to be significantly smaller than in the chaotic regime. In real cells, where the regime is believed to lie on the edge of 
chaos, the cascades tend to be small also. Moreover, the distribution of the avalanche sizes in the ordered regime follows a power-law curve \cite{kauffman2000}, with many small and few large avalanches. In the chaotic 
regime, in addition to the power-law distribution, 30-50 percent of the avalanches are huge. The distribution of avalanche sizes of RBNs in the ordered regime roughly fits the expectations of biologists, where most of the 
genes, if perturbed, are only capable of initiating a very small avalanche, if any. Fewer genes could cause bigger cascades, and only a handful can unleash massive ones.
Perturbing an arbitrary gene is reasonable in RBNs where all genes have the same average number
of interactions. In scale-free nets however, this is no longer true due to the presence of a high
degree inhomogeneity. Even for values of $\gamma$ around $3.5$ there will be nodes that have many
more output connections than the average value. A transient perturbation of a gene that has few
interactions will have moderate or no effect, while perturbing a highly connected node will have
larger consequences.
Several studies dealing with various kinds of system perturbation have been recently published.
Aldana's approach~\cite{aldana-sf-03} is similar to the one taken here except that he deals
with small scale-free networks in which $N$, the number of nodes, is $19$. Ribeiro and 
Kauffman~\cite{rib-kauff-07} exhaustively studied the state space of small ($N < 20$) RBNs
under probabilistic errors in gene state searching for ergodic sets, i.e. sets of states such that
once the system is in one of them, it cannot leave it subject to internal noise. They find that if noise
may affect all nodes of an attractor then multiple ergodic sets are unlikely. However, when noise
is limited, multiple ergodic sets do exist which means that attractors are stable. Serra et al~\cite{serra-jtb-07} present a study of the distribution of avalanches in unperturbed RBNs and in RBNs in
which one gene has been ``knocked-out'', i.e. a state $0$ has been permanently changed to $1$. They show that
the standard model readily explains the distribution of the resulting avalanches. They also
examined the influence of a scale-free topology for the outgoing links on the system. Aldana
et al.~\cite{aldana-faults-07} examine the effect of more complex and biologically plausible perturbations
of the attractor landscape of both standard RBNs and scale-free RBNs. Genes undergo duplication
and mutation which cause topological changes that in general maintain the original attractors
and may create new ones. Near the critical regime robustness and evolvability are found
to be maximum.

\subsection{The Effect of Perturbation}  

In this work we have submitted all systems that have reached biologically plausible attractors to ``gene damage'', the simplest failure amongst those previously described. That is, when the system is cycling through the configurations of the attractor, the whole system is duplicated. The original will 
continue unperturbed. On the other hand,  a node of the copy is chosen at random and will give the opposite output value a single time step. This usually knocks the system out of the course of its attractor. Now we let both 
systems evolve over time and record at each time step how many more nodes have a different value in the copy compared to the original. This value usually reaches a maximum that represents the number of nodes that have 
ever had a behavior different than those of the original. This number is the size of the {\it avalanche}. There are only three possible senarios for the copy: it will return to the same attractor as the original, it will reach a different 
attractor or diverge and reach no attractor within the maximum number of configurations allowed (1000). Each system in an attractor is copied 10 times, and each copy will have a different avalanche starting point. We record 
separately these informations in order to compare the re-convergence capabilities of systems in each regime, with different topologies and update schemes. \\
Fig. \ref{attRecov} shows the frequency at which systems that have already converged to an attractor re-converge. We show separately whether systems re-converge to any attractor or to the same one as before the perturbation. In particular, Fig. \ref{attRecov} depicts results for attractors before perturbation (original attractors) of sizes between 2 and 100. We show systems that used networks of size $N=200$. Results for smaller systems are comparable and are not shown in this work.
\begin{figure*} [!ht]
\begin{center}
\begin{tabular}{cc}
\mbox{\includegraphics[width=6.5cm]{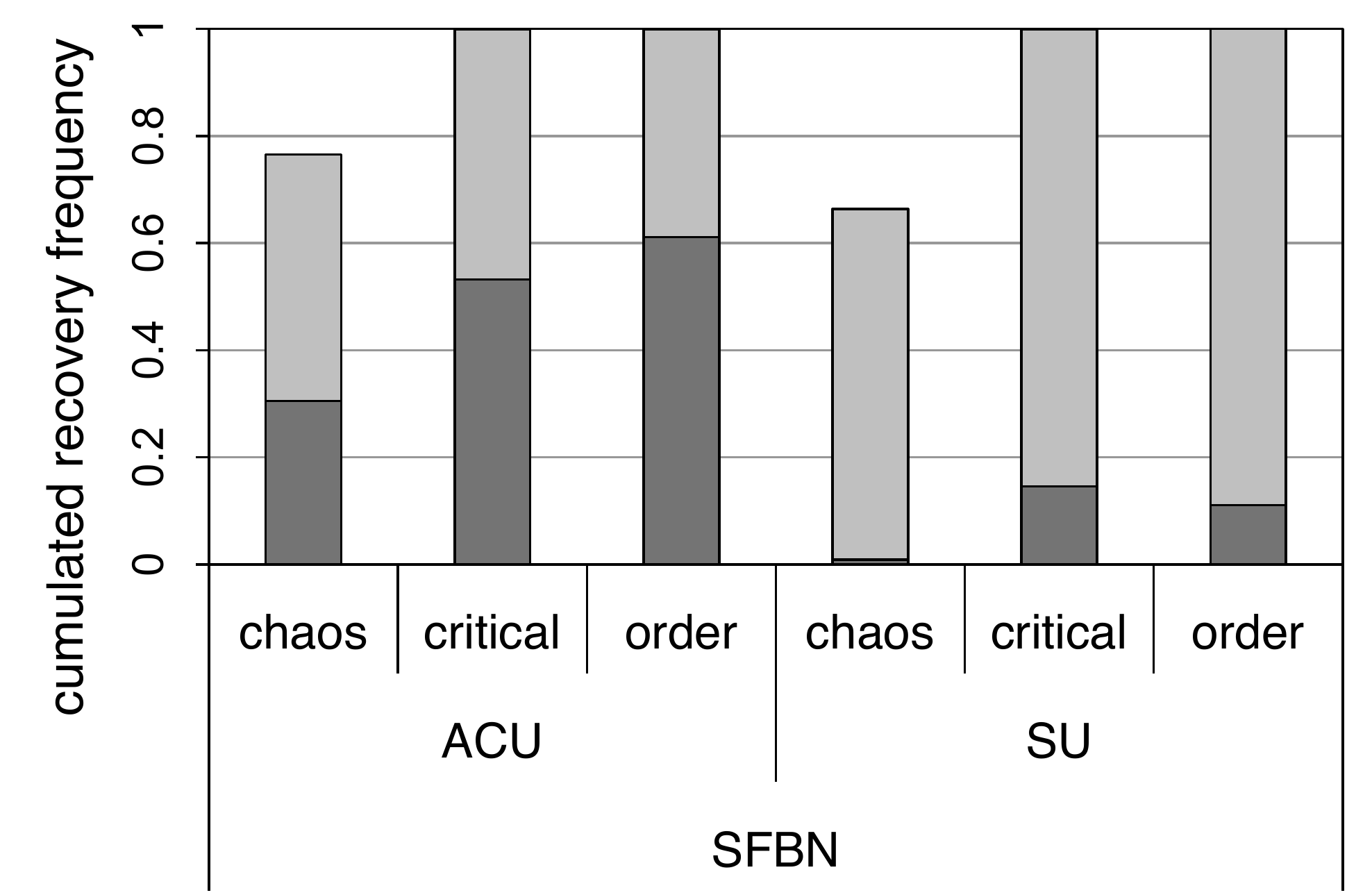} } \protect &
\mbox{\includegraphics[width=6.5cm]{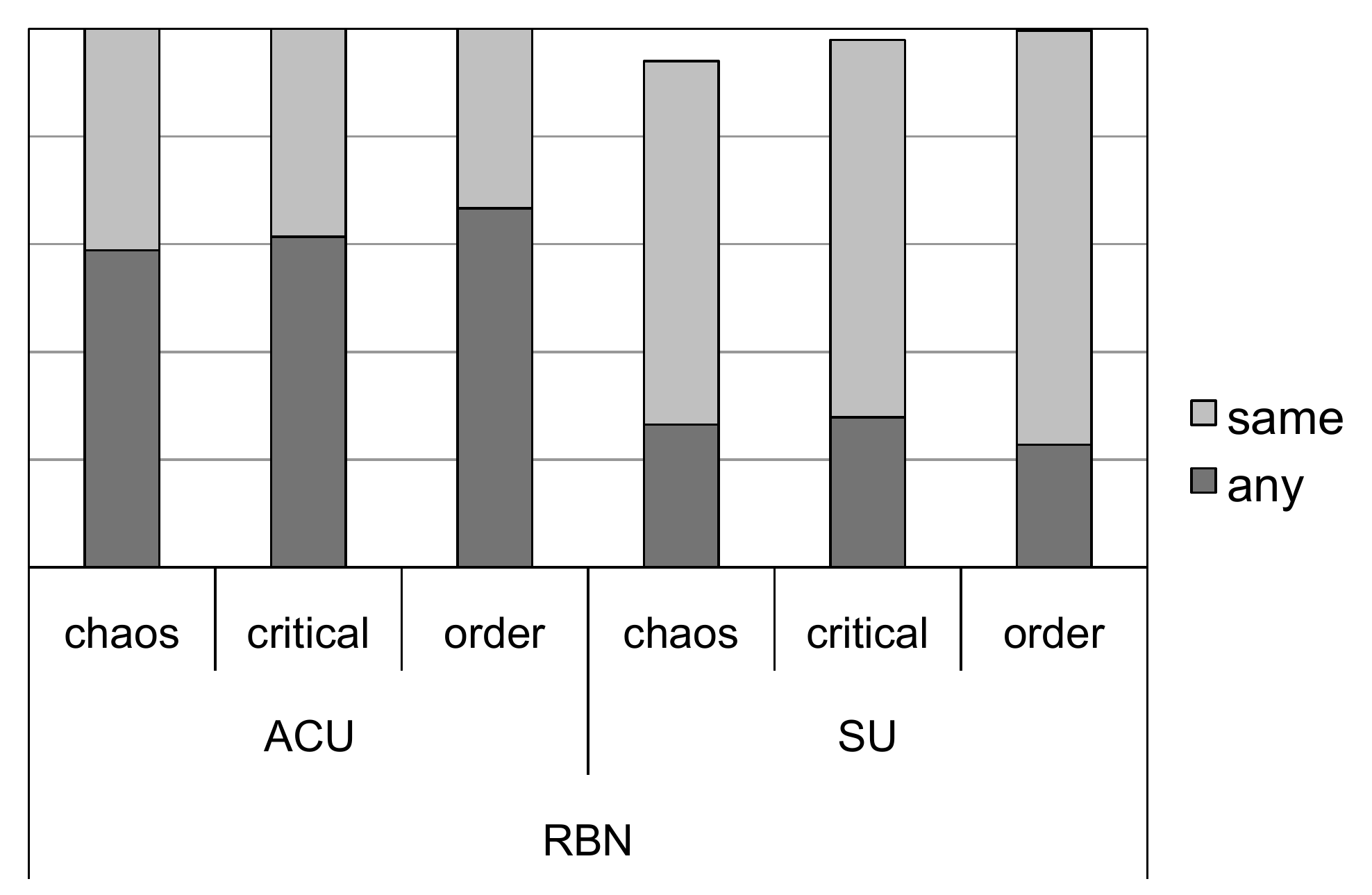} } \protect \\
(a)  &  (b) \\
\end{tabular}
\end{center}
\caption{Re-convergence frequency. Frequency at which perturbed systems re-converge to either the same attractor (light grey) or another one (dark grey). Left-hand (a) side figure shows results for SFBNs and right-hand (b) side figure shows results for RBNs. All systems have $N=200$ nodes. We purposefully omit point attractors.}
\label{attRecov}
\end{figure*}
Re-convergence seems to mostly depend on the regime the system evolves in, rather than its degree distributions, update scheme, or size. On Fig. \ref{attRecov} we note that only networks in the chaotic regime do not re-converge to an attractor in every case. It also seems that ACU performs a little better at helping systems to find a stable state. However, this tendency seems inverted when taking into account only cases where the same attractors are found. In this case, under ACU, the same attractor as the original one is found about half of the time. Under SU, the same one is found about 75\% of the time. This could be explained by the fact that the number of attractor lying in the state spaces of systems under ACU is much greater.\\
As expected when dealing with random failure, the information traveling through a structure with regular output distribution is more vulnerable to faults compared to structures with {\it hubs} and {\it leaves}. This fact is well known in various examples such as computer networks which are very resistant to random failure as long as they are failures and not targeted attacks on highly interconnected nodes. Especially under SU, SFBNs tend to re-converge to the same attractor more than RBNs, although overall, both topologies perform well. The chaotic case will be explained below in details. Under ACU, critical and ordered SFBNs systems are again performing as able as or better than their counter parts in RBNs, recovering as often to any attractor but more often to the same as the original one. The counter-performance of chaotic systems, especially SFBNs, can be explained by the ``spike of huge avalanches'' described by Kauffman \cite{kauffman2000} and visible in Fig. \ref{avalancheSize}. Indeed,  SFBN systems and, in a lesser manner, RBN under SU have a surge of very long avalanches when in the chaotic regime. This characteristic explains why these systems are not as well able to re-converge to an attractor, let alone the same one.\\
Fig. \ref{avalancheSize} shows the distribution of the avalanches' size. Again we distinguish networks that have re-converged at all in Fig. \ref{avalancheSize}(a) and those that have re-converged to the original attractor Fig. \ref{avalancheSize}(b). For readability reasons and, since results are very similar, we show only results for systems of size $N=100$ and $N=200$.
\begin{figure*} [!ht]
\begin{center}
\begin{tabular}{ccc}
& any attractor & same attractor\\

\raisebox{10ex}[0pt]{\begin{sideways}$N=100$\end{sideways}} 
&\mbox{\includegraphics[width=6cm]{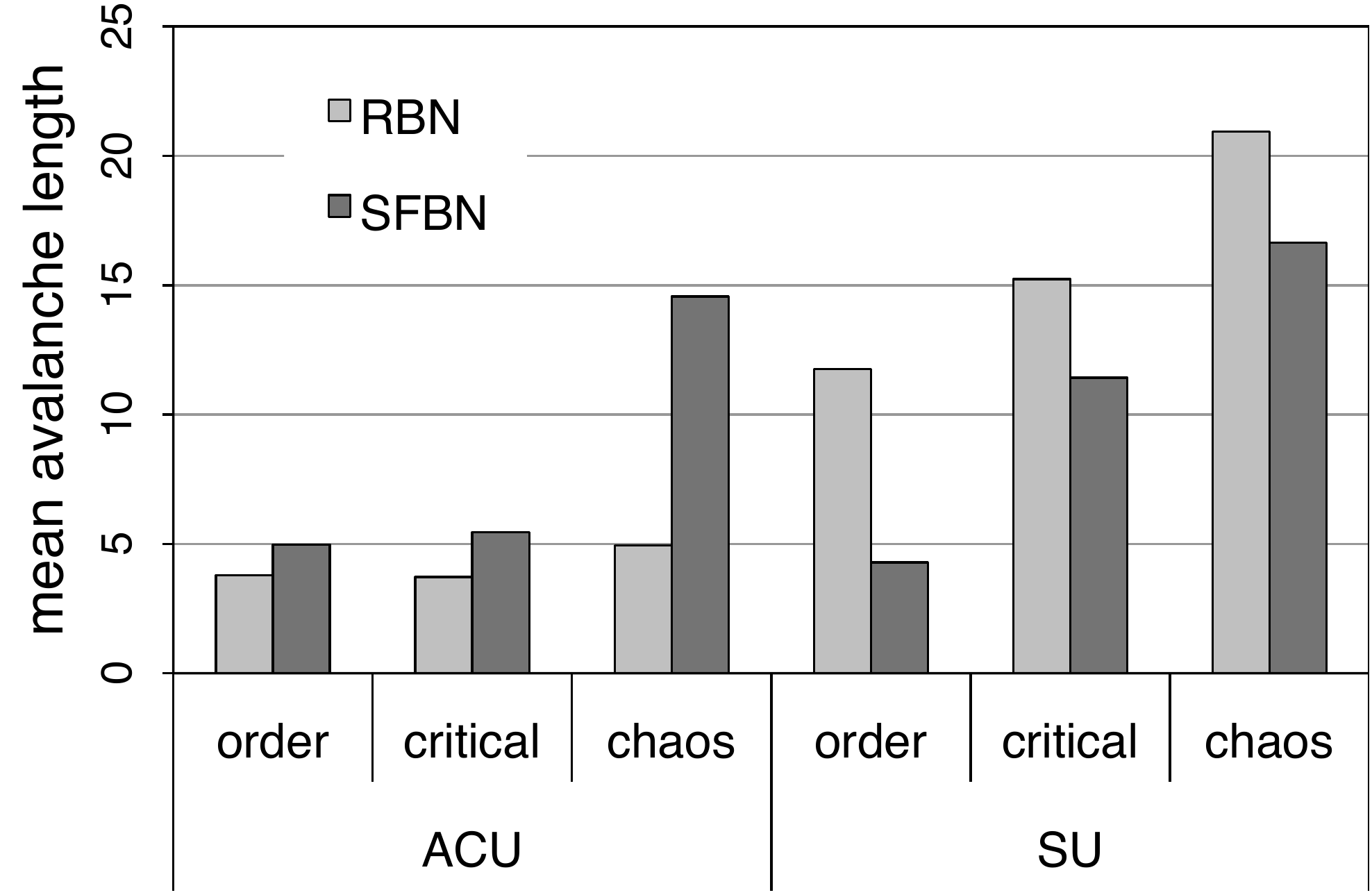} } \protect 
&
\mbox{\includegraphics[width=6cm]{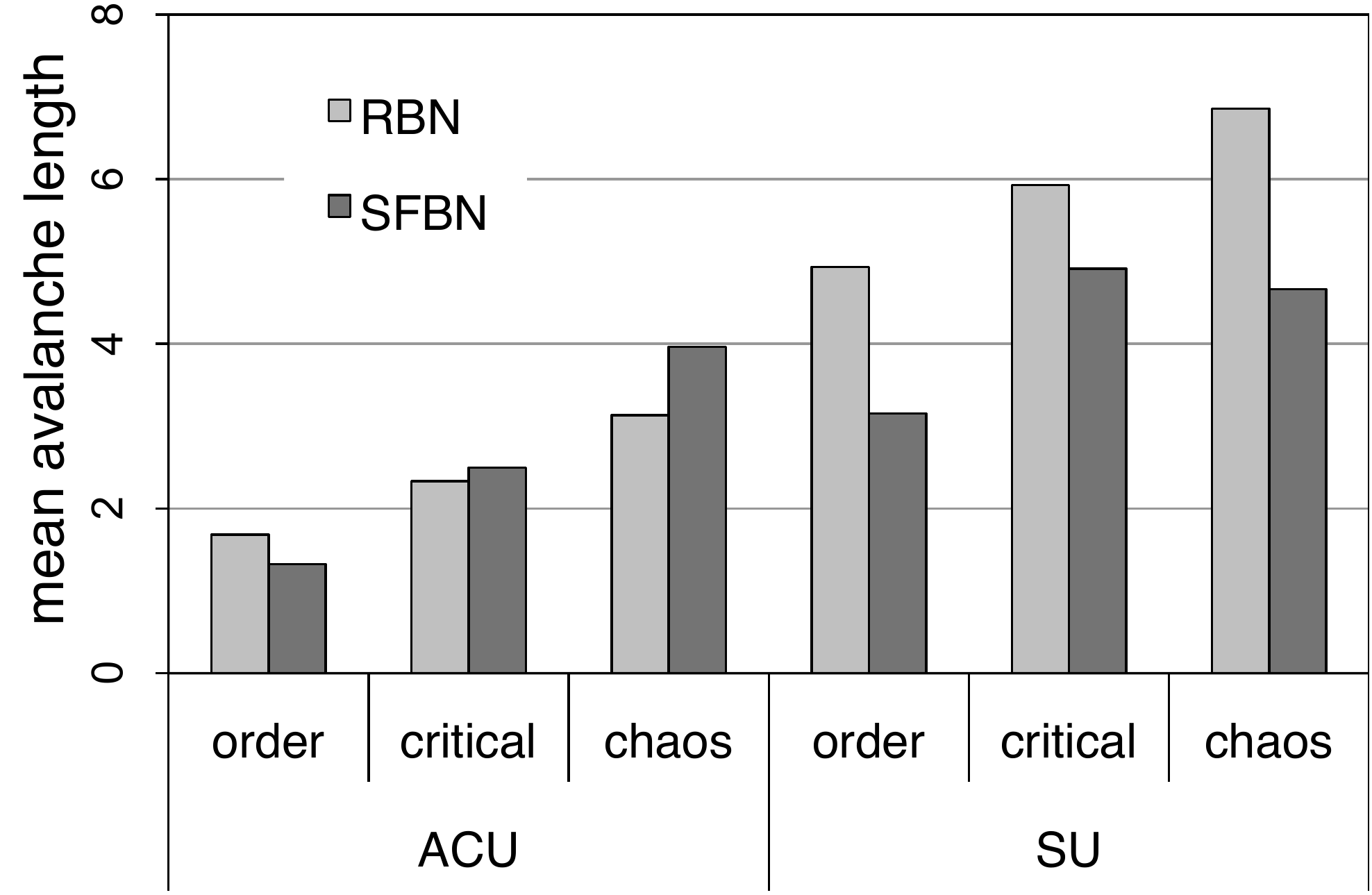} } \protect \\
&(a)  &  (b) \\

\raisebox{10ex}[0pt]{\begin{sideways}$N=200$\end{sideways}} 
&\mbox{\includegraphics[width=6cm]{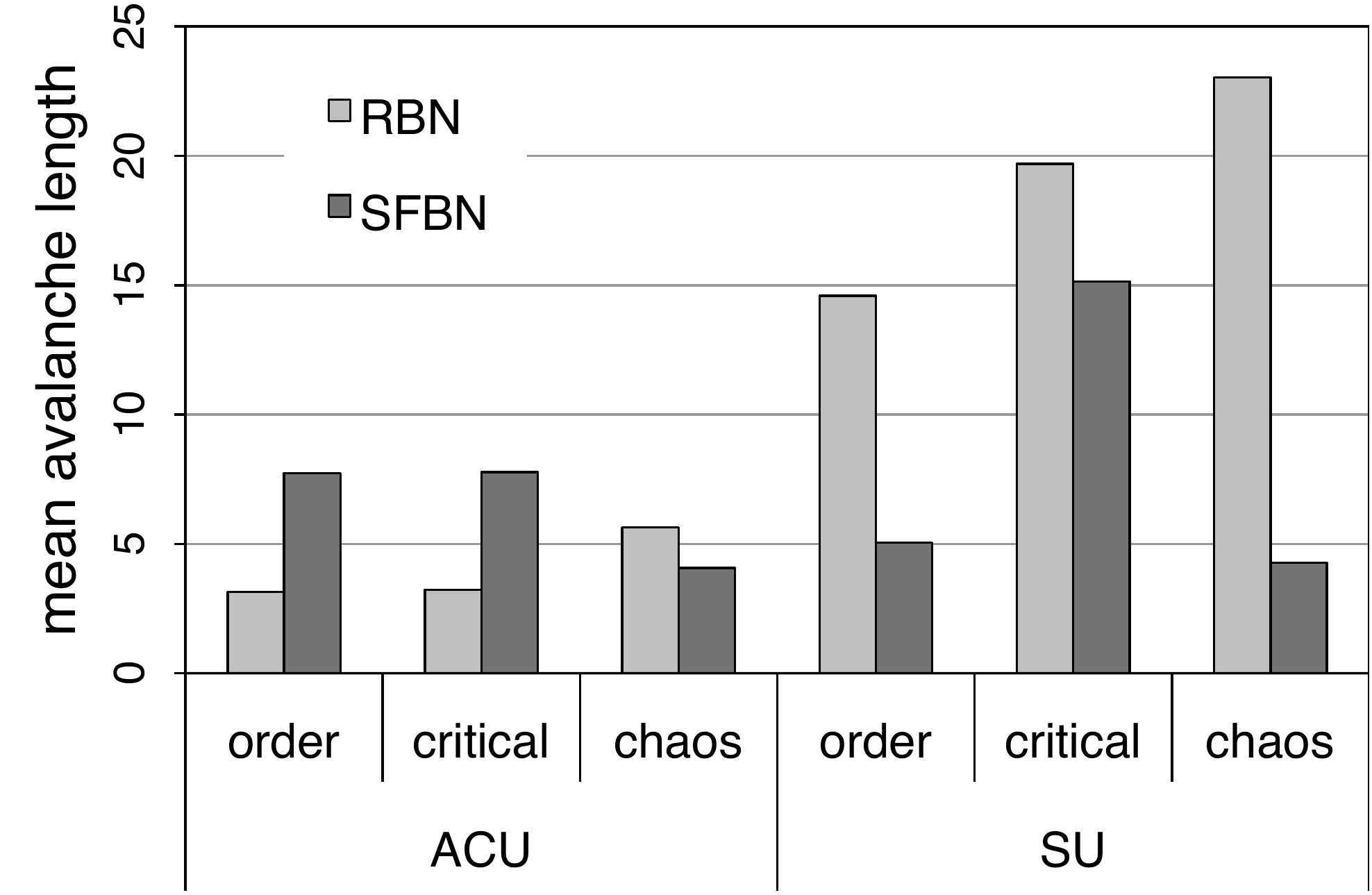} } \protect 
&
\mbox{\includegraphics[width=6cm]{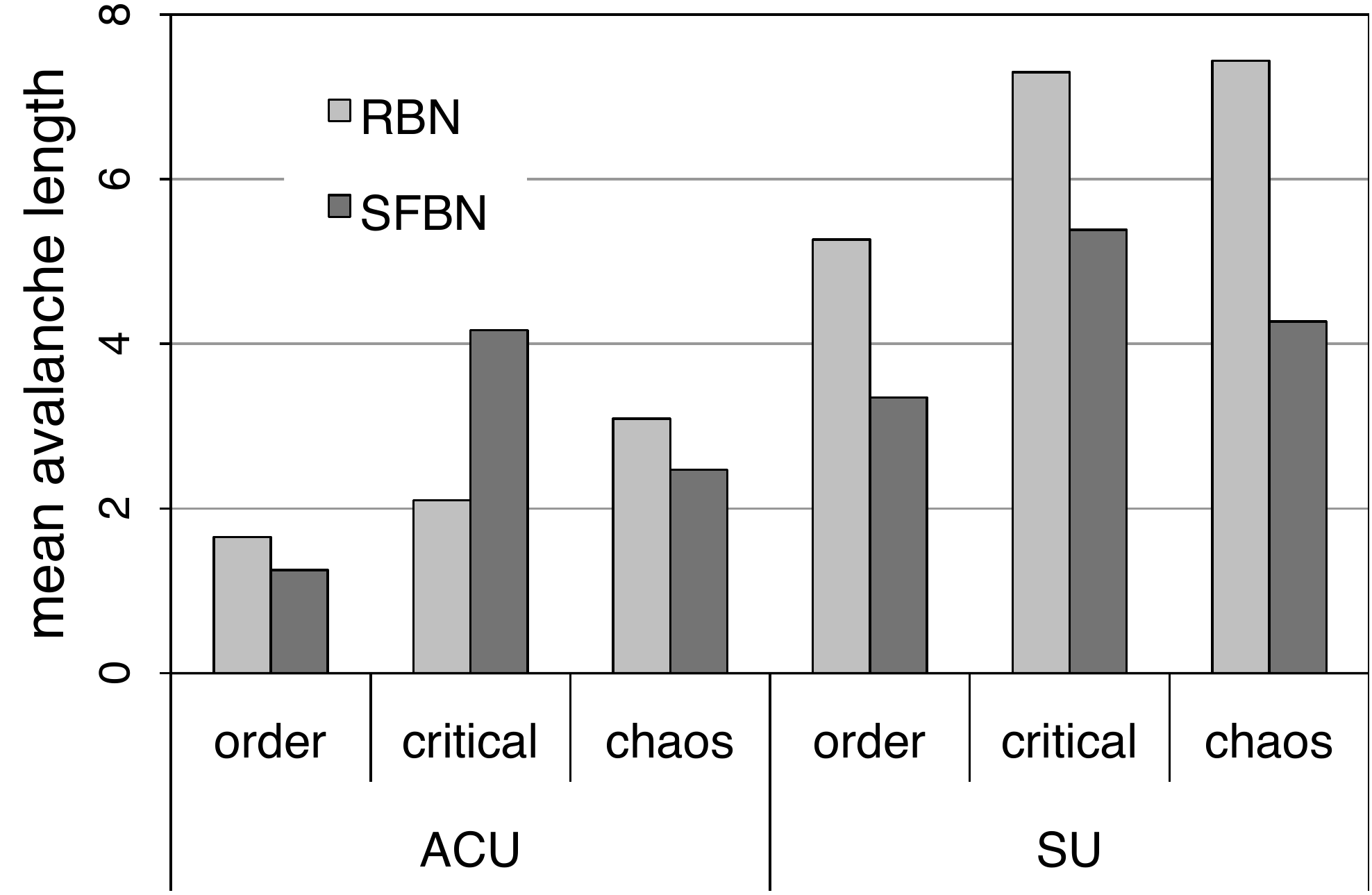} } \protect \\
&(c)  &  (d) \\
\end{tabular}
\end{center}
\caption{Mean avalanche lengths. Average avalanches length of cases where systems re-converge to (a)(c) any attractor and (b)(d) the same attractor.}
\label{avalancheSize}
\end{figure*}
As mentioned in Section \ref{sect:fault}, the size of the avalanche varies mainly due to the regime. Smaller systems with $N=100$ react as expected, with the size of their avalanches increasing as the systems grows chaotic. 
However, this does not seem to always be the case, and this relationship between avalanche size and regime is changed in bigger networks. Under ACU networks where $N=150$ or $N=200$, it is the systems that evolve in the critical regime that clearly show the longest avalanches. This is true for ACU only, SU systems still corroborate Kauffman's conjecture. Although in the case where systems return to the original attractor, avalanche sizes are much smaller, the tendencies observed in the more general case stand. This time we observe an obvious impact of the networks size on the systems dynamics. Further investigations are necessary as to define why larger systems in critical regime under ACU are more impacted by perturbations.\\
Fig. \ref{avalDistr} shows the distribution of the avalanches' sizes for different systems. Although values are discrete, we used continuous lines as a guide for the eye.
\begin{figure*} [!ht]
\begin{center}
\begin{tabular}{ccc}
& ACU & SU\\

\raisebox{10ex}[0pt]{\begin{sideways}SFBN\end{sideways}} 
&\mbox{\includegraphics[width=6cm]{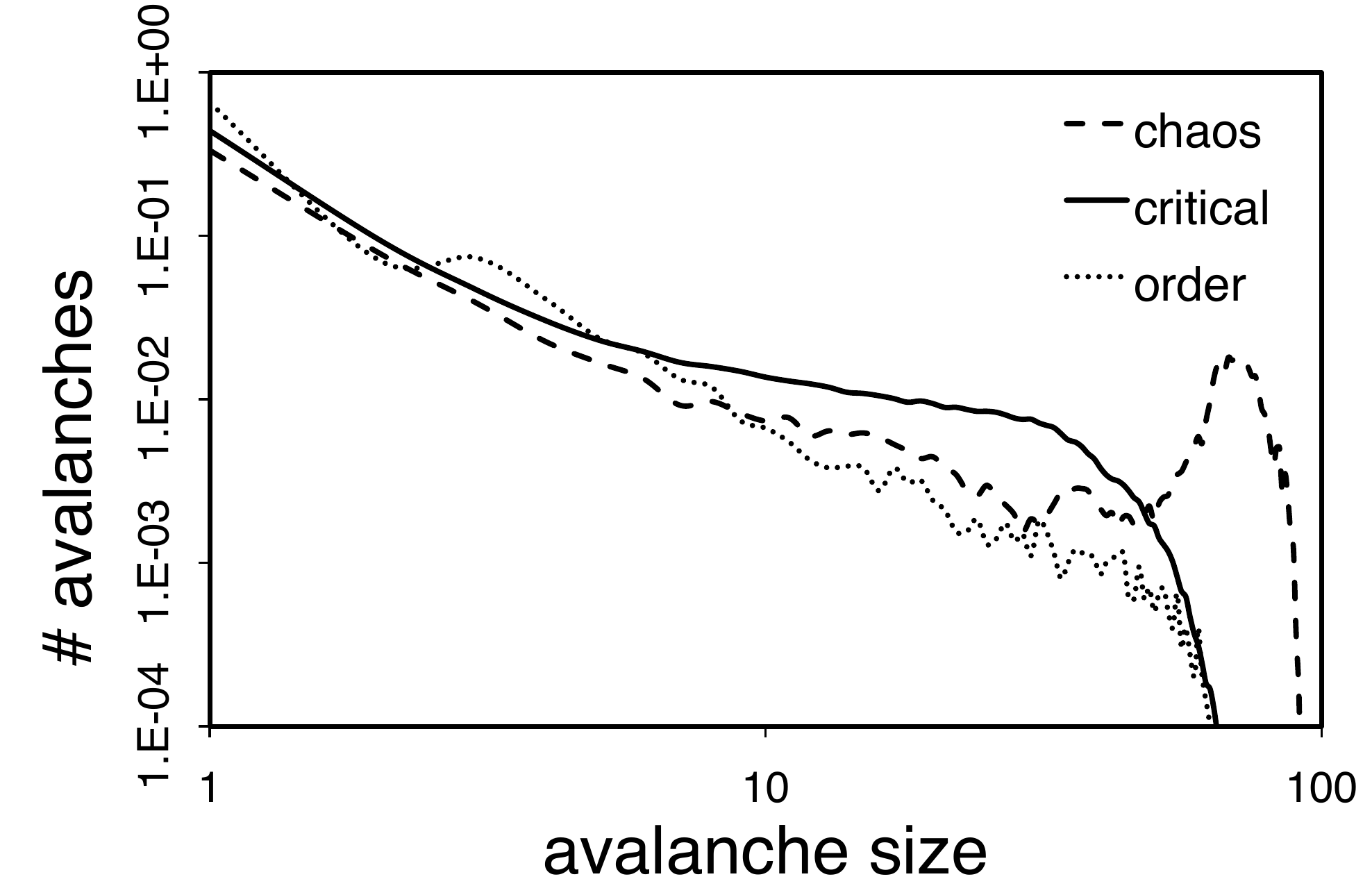} } \protect 
&
\mbox{\includegraphics[width=6cm]{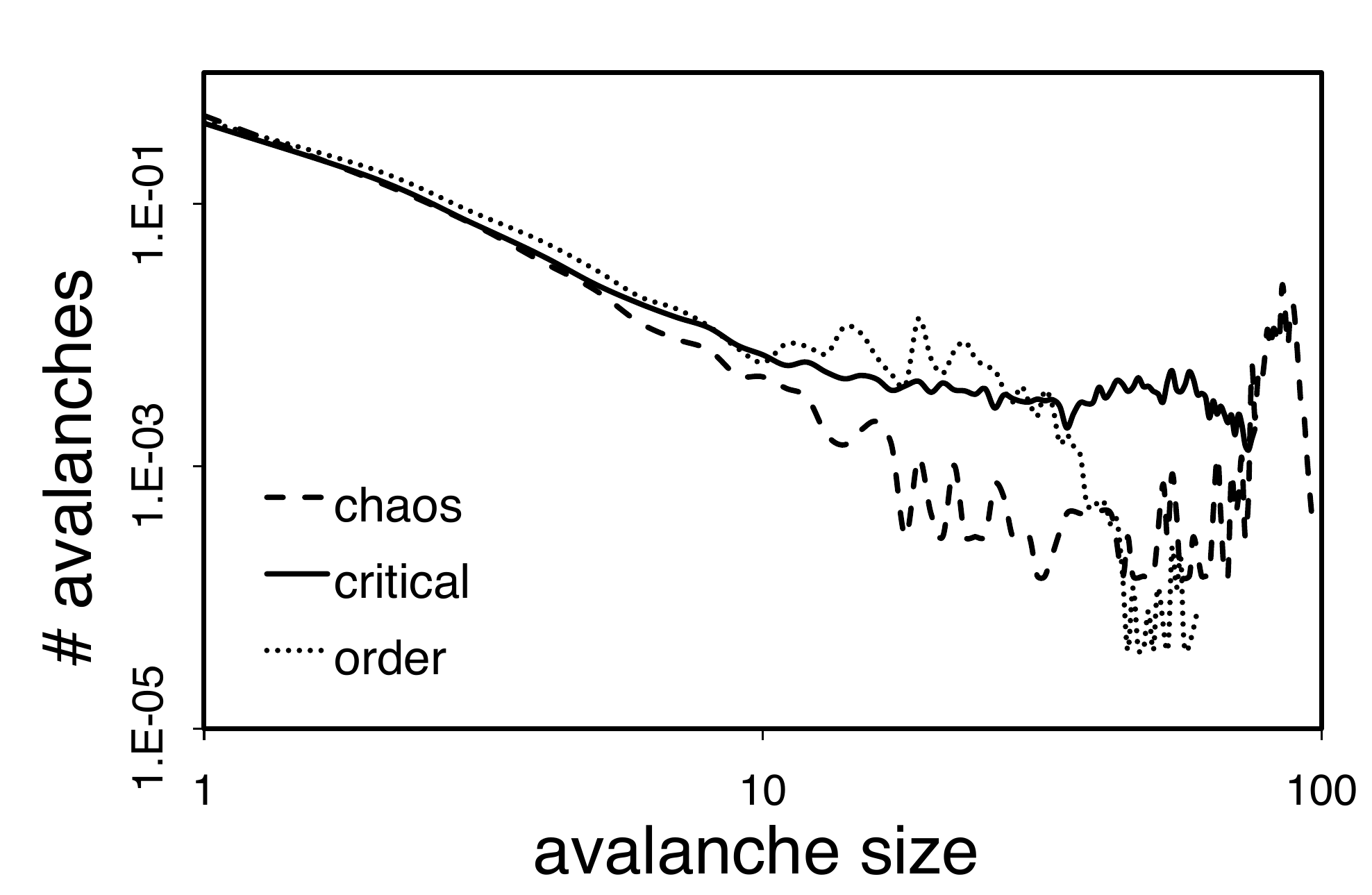} } \protect \\
&(a)  &  (b) \\

\raisebox{10ex}[0pt]{\begin{sideways}RBN\end{sideways}} 
&\mbox{\includegraphics[width=6cm]{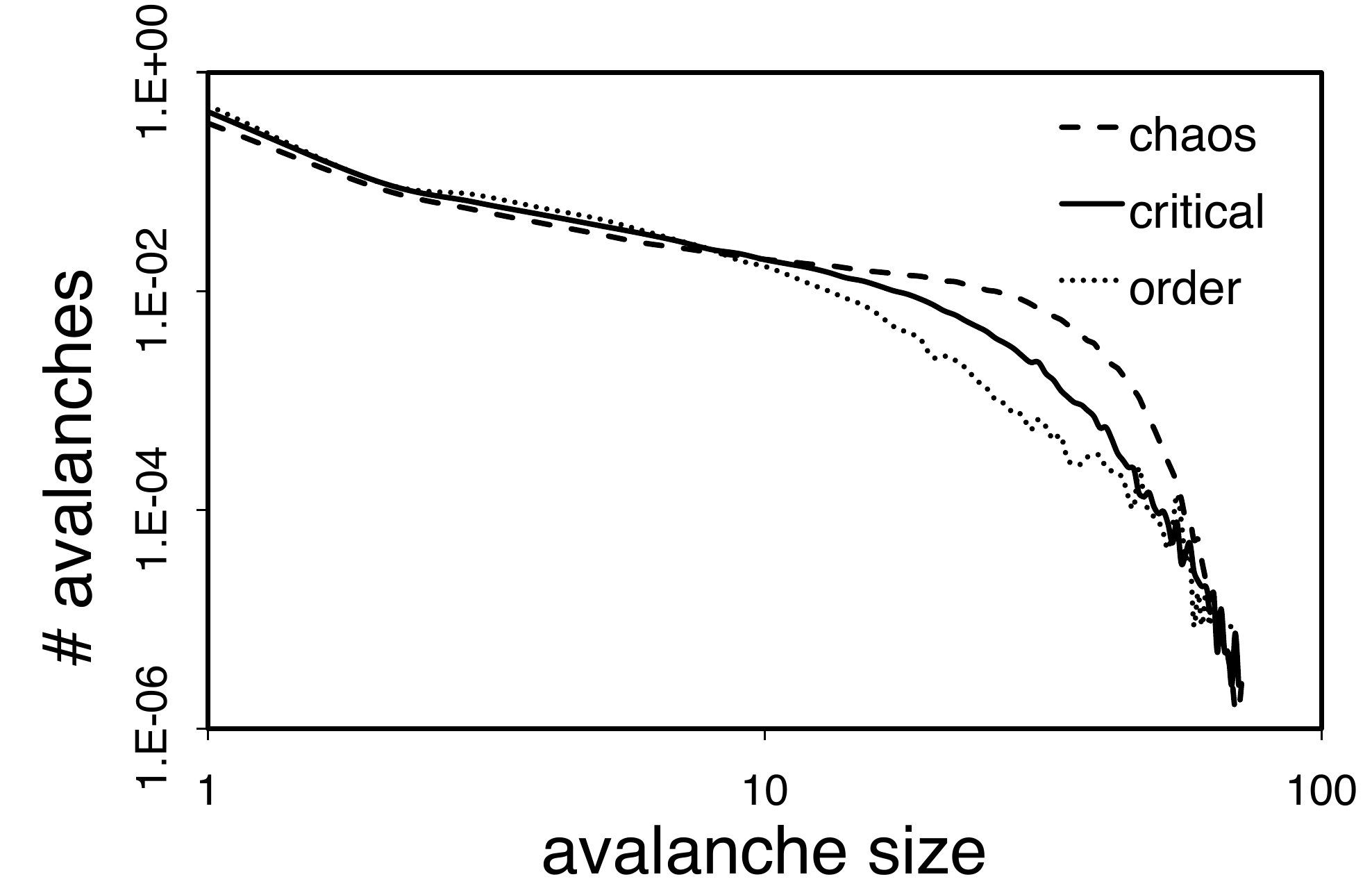} } \protect 
&
\mbox{\includegraphics[width=6cm]{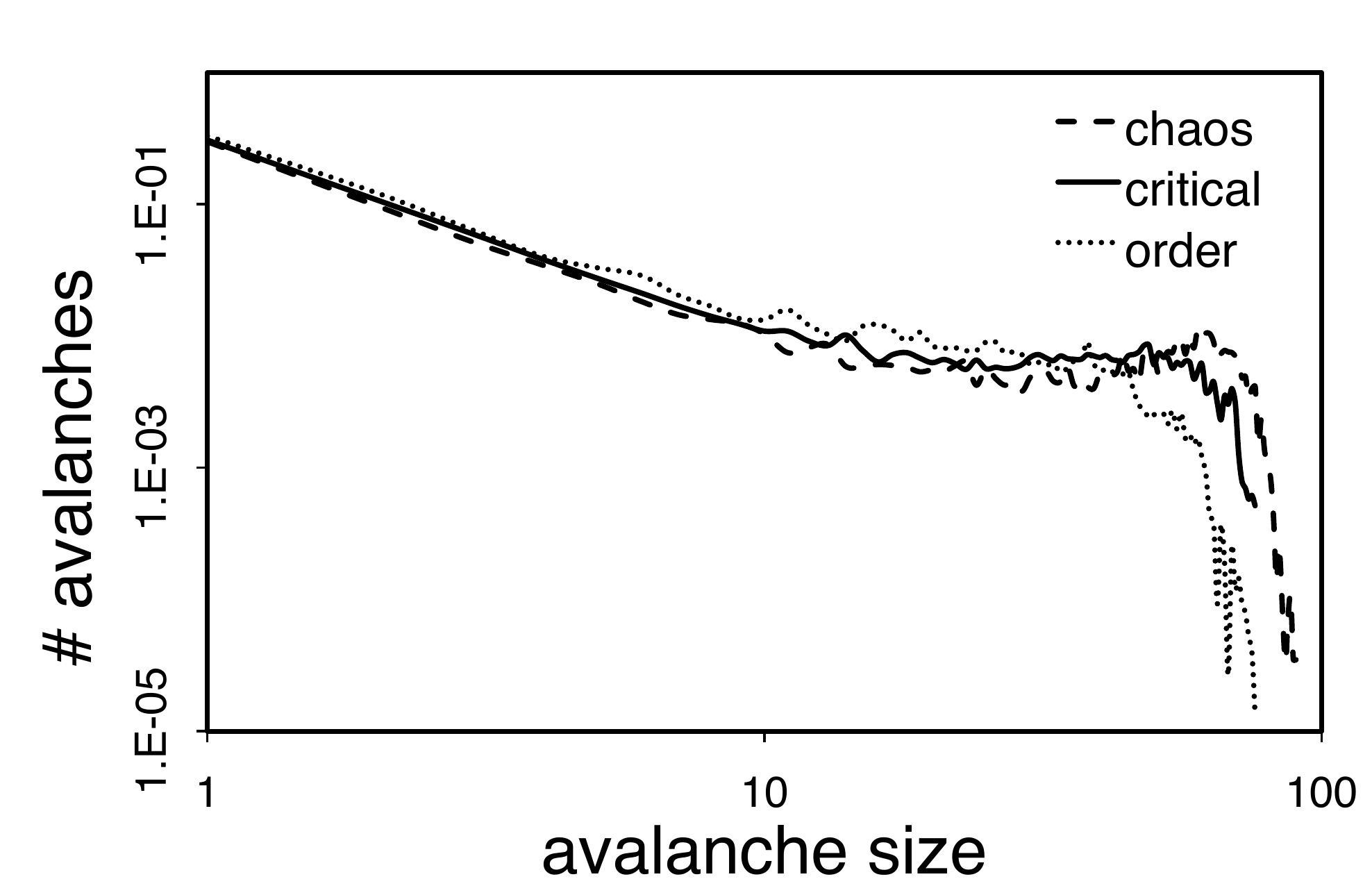} } \protect \\
&(c)  &  (d) \\
\end{tabular}
\end{center}
\caption{Avalanche length distributions. Distribution of the avalanche lengths for all three regimes and $N=100$. 
Upper row (a)(b): For SFBNs. Lower row (c)(d): For RBNs. Left-hand side column (a)(c): Systems under ACU. Right-hand side column (b)(d): Systems under SU.}
\label{avalDistr}
\end{figure*}
In Fig. \ref{avalDistr}, we see that tendencies are the same and are anticipated from Kauffman's work \cite{kauffman2000}. SFBNs under both (a) SU and (b) ACU exhibit a steady long tailed decrease in the number of avalanches as their length grows for ordered and critical regime, and there is an increase for long avalanches in the case of chaotic systems. This tendency is the same for synchronous RBNs in (d). Interestingly, this does not seem to apply to RBNs under ACU, where no increment is to be noted. \\
Lastly, Fig. \ref{avalAvgDegree} illustrates the average output degree of the node that represents the damaged gene. For clarity reasons, we show results only for bigger systems as they are similar when networks are scaled 
down.
\begin{figure*} [!ht]
\begin{center}
\begin{tabular}{cc}
\mbox{\includegraphics[width=6.5cm]{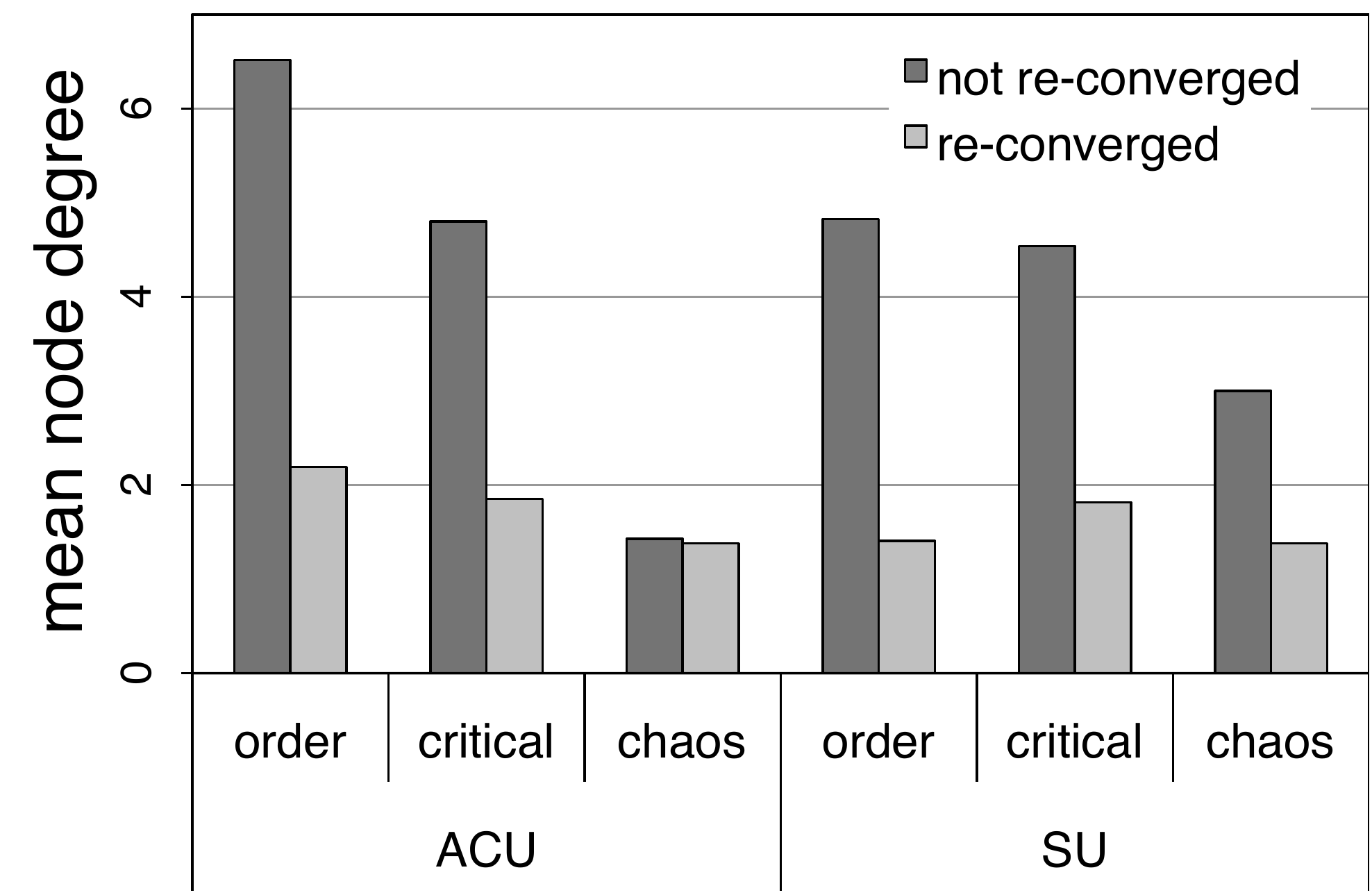} } \protect
&
\mbox{\includegraphics[width=6.5cm]{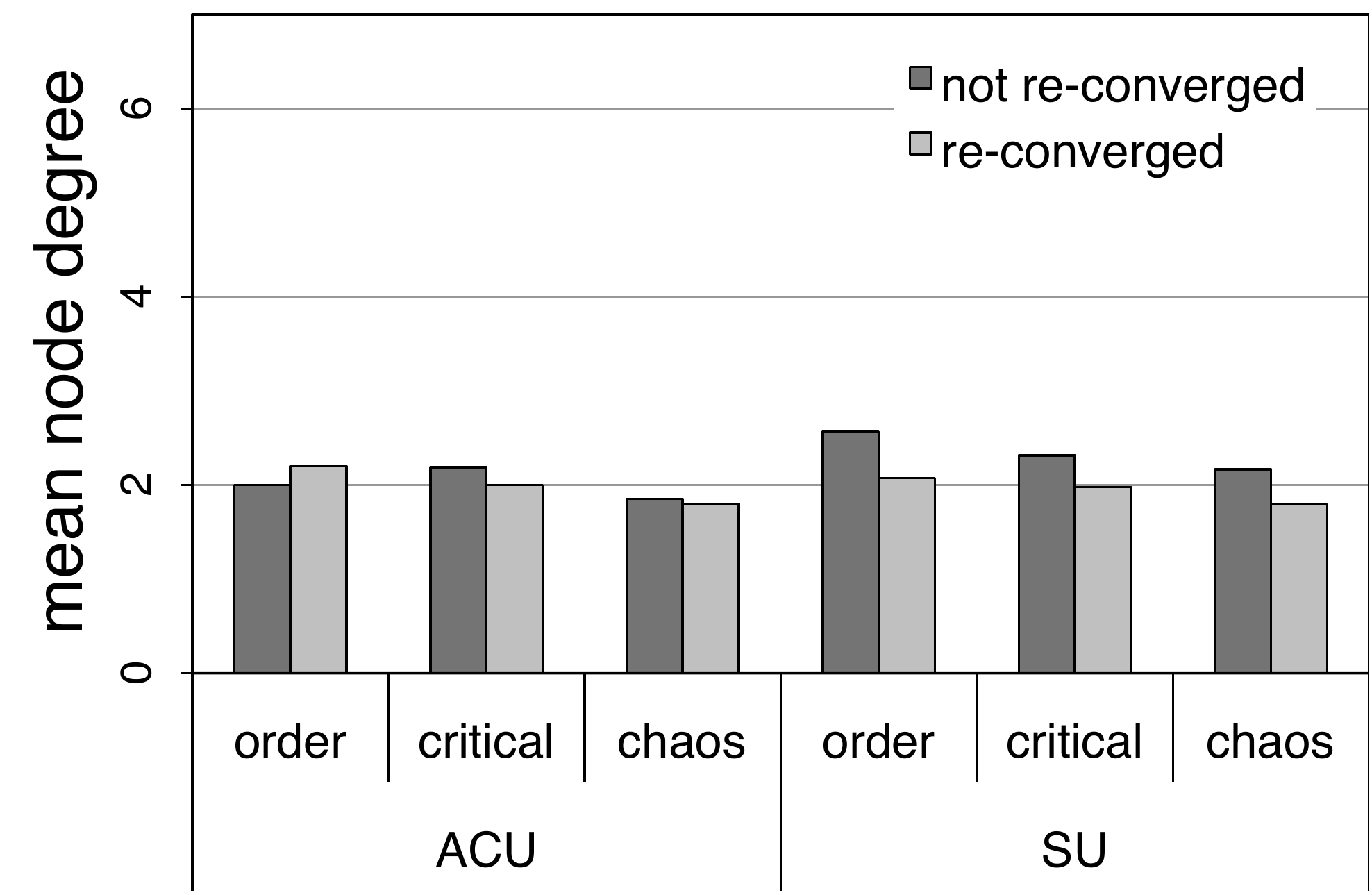} } \protect \\
(a)&(b)
\end{tabular}
\end{center}
\caption{Faulty nodes average degree. Average degree of the nodes that have failed for both re-converged and not re-converged avalanches. RBNs are shown on the right hand side (b) and SFBNs left one (a). Size $N=200$.}
\label{avalAvgDegree}
\end{figure*}
Although predictable, we clearly see the effect of the hubs in SFBNs, where failing nodes in systems that do not re-converge had a much higher output degree on average than those of systems that did recover. Another interesting observation, is that there seems to be a direct relationship between the degree of the wrongful node and the regime, the more ordered the system, the higher the degree to allow the system to recover. This difference is toned down in ACU systems. Naturally this does not hold for classical RBNs, where all nodes have 
comparable output degrees.\\
As a general conclusion on gene-damage failures of Boolean Networks, we can highlight the prominent effect of the topology on distribution of the lengths of the avalanches and its ability to re-converge to an attractor over the networks update and regime.

\subsection{Derrida Plots} 

In this section we compare Derrida plots of our models with those of Kauffman~\cite{kauffman2000} and Iguchi 
et al.~\cite{iguchi07}. These representations are meant to illustrate a convergence versus a divergence in state space that can in turn help characterize the different regimes. These plots show the average Hamming distance~\footnote{The normalized number of positions that are not identical when comparing two (binary) strings.} $H(t)$ between any two states $s_a$ and $s_b$ and the Hamming distance $H(t+1)$ of their respective consecutive state $s_a'$ and $s_b'$ at the next time step. Derrida plots of systems in the chaotic regime will remain above the main diagonal $H(t)=H(t+1)$ longer, crossing the main diagonal earlier and remaining closer to it as the systems near the critical regime. Systems in the critical regime remain on the main diagonal before diverging beneath it. Ordered systems remain under the main diagonal at all times.  These results are
already known for RBNs under SU and, to some extent as the regimes are not defined explicitly by Iguchi et al.~\cite{iguchi07}, for SFBNs under SU. \\
Figure~\ref{derrida} shows the Derrida plots for systems under ACU on the right and SU on the left. This plot concerns networks of size $N=100$.
\begin{figure*} [!ht]
\begin{center}
\begin{tabular}{cc}
\mbox{\includegraphics[width=6.5cm]{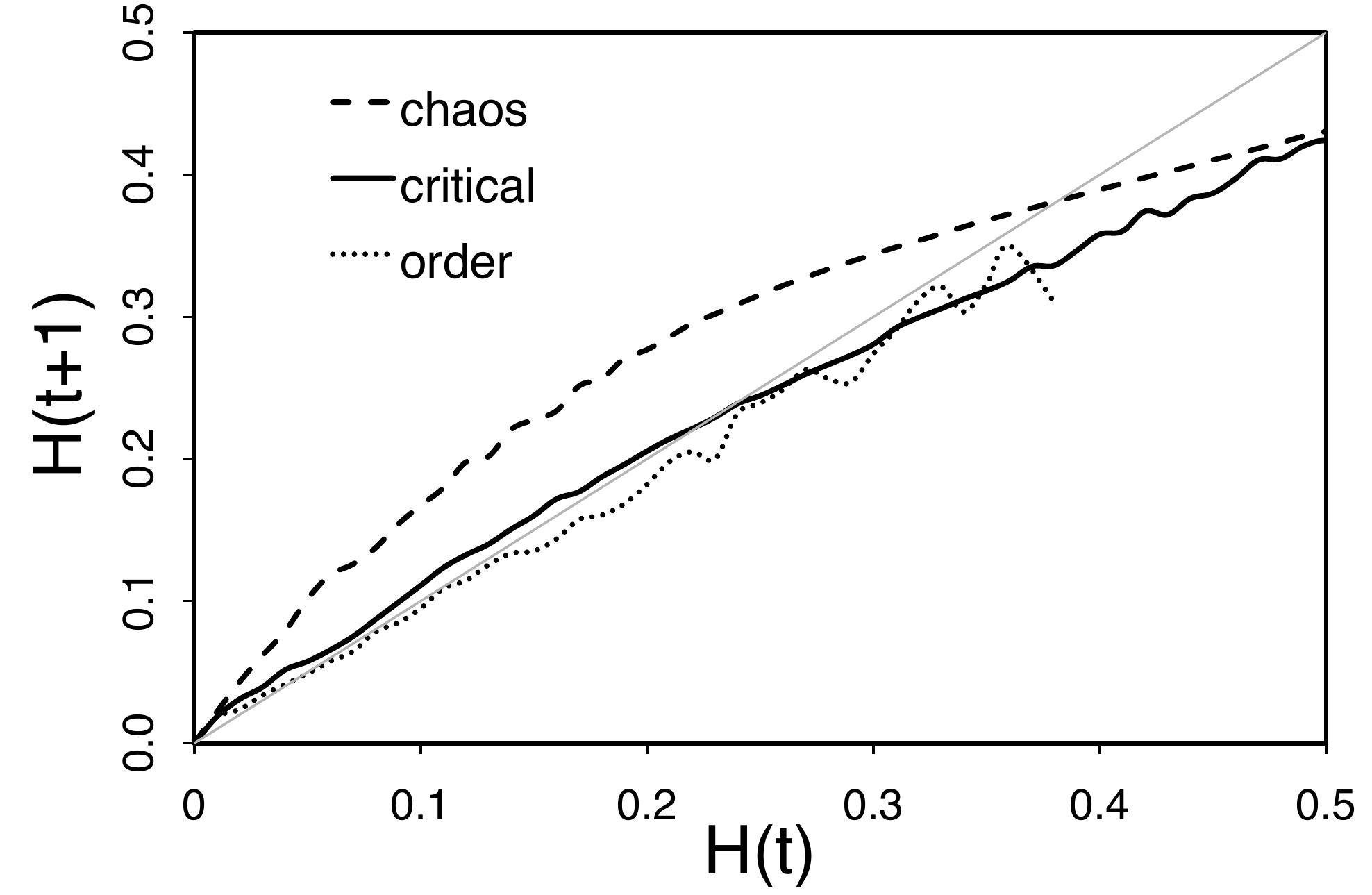} } \protect 
&
\mbox{\includegraphics[width=6.5cm]{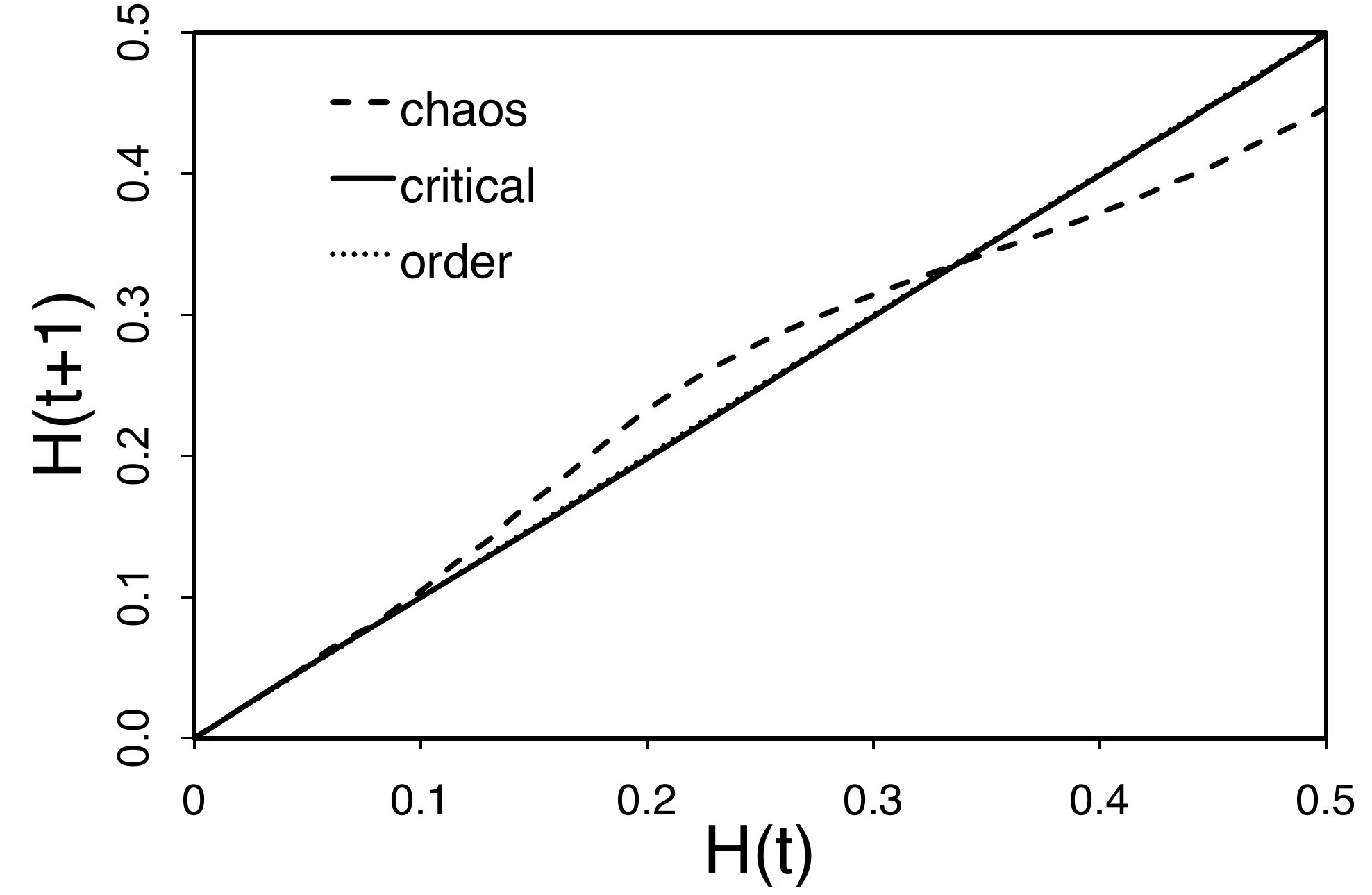} } \protect \\
SU & ACU \\
\end{tabular}
\end{center}
\caption{Derrida plot of the Hamming Distance $H$ at time $t$ vs. $t+1$. The right-hand side figure represent systems under Activated Cascade Update and the left-hand side one under Synchronous Update. All systems are SFBNs of size $N=100$. }
\label{derrida}
\end{figure*}
The system under SU on the right-hand side has a fairly typical behavior, where the chaotic systems remain clearly above the main diagonal, critical ones remain close to and then diverge below the main diagonal. The ordered curve, although clearly remaining below the main diagonal, is somewhat irregular. This is probably due to the fact that the number of attractors is the lowest of all in the ordered SFBN systems under SU, thus making the curve less smooth. In the case of SFBNs under ACU, the chaotic curve shows the expected behavior, though it remain closer to and crosses the main diagonal earlier than in the SU case. It can still be considered a reliable indication that the system is indeed in a chaotic regime. In the ordered and critical regimes, the curves are literally on the main diagonal all the way through, and show no sign of convergence or divergence whatsoever. So in this case, the update method has a major impact on the Derrida plots, making the ordered and critical systems impossible to distinguish under ACU. 


\section{Conclusions and Future Work}
\label{sect:concl}

Although a long way from a fully functional model of GRNs, we are moving closer to one by aggregating modern findings obtained with recent high throughput techniques. These refinements to the original RBN model by Kauffman and the subsequent ones by Aldana help us understand some key details of the complex interactions that are taking place between the different components and the role that the topological structure plays in the dynamics. In this paper, we have made some
progress towards an understanding of what structural and dynamical properties make GRNs highly stable  and adaptable to mutation,
yet resistant to perturbation.\\
This work suggests one structural property, namely the scale-free output distribution, and a dynamical one, the semi-synchronous updating, to try to improve the standard RBN model and to account in an abstract way for recent findings in system-level biology. We have used computer simulations to reflect the impact of these changes on original RBN models. Results are encouraging, as our SFBNs model shows comparable or better performance than the original one with more attractors and smaller avalanches. This leads us to believe that the models are pointing in the right direction.
Nevertheless, from the results of this analysis, we also see that neither model is the absolute optimum in this problem. Indeed, if we focus on maximizing the number of attractors, the prominent effect is that of the update, with ACU combined with original RBNs achieving the best results in finding the most attractors with a biologically relevant cycle length. On the other hand, when considering maximizing the fault tolerance, we witness the highest resilience with SFBNs under SU, that achieve the highest rate of re-converging to the same attractor as observed originally. This demonstrates that no combinaison is optimal on all problems and that compromise is necessary if we are looking to build a model that will perform well in a realistic situation.\\
In the future, we intend to expand the range of analysis conducted on perturbed systems, in the hope of shedding some light on GRNs. Also, we would like to explore different degree distribution types and combinations, including the use of
some actual GRNs as high-throughput molecular genetics methods make real-life data available like never before.


\subsubsection*{Acknowledgements}

The authors thank F. Di Cunto and P. Provero of the University of Torino (Italy) for the 
useful discussions and suggestions on biological regulatory networks and C. Damiani, M. Villani and R. Serra for their insightful suggestions about RBN models. M. Tomassini and Ch.~Darabos 
gratefully acknowledge financial support by the Swiss National Science Foundation under contract 
200021-107419/1. M. Giacobini acknowledge funding (60\% grant) by the Ministero dell'Universit\`a e della Ricerca Scientifica e Tecnologica.

\bibliographystyle{elsart-num}
\bibliography{jtb08}

\end{document}